\documentclass[11pt,preprint]{article}
\usepackage{amssymb}
\usepackage{color}
\usepackage{amsmath}
\usepackage{graphicx}
\usepackage{multirow, multicol}
\usepackage{rotating}
\usepackage{hyperref} 

\usepackage{amssymb}
\usepackage{amsmath}
\usepackage{graphicx}
\usepackage{hep}
\usepackage{soul}
\usepackage{latexsym}
\usepackage{amsfonts}
\usepackage[latin1]{inputenc}
\usepackage{graphicx}
\usepackage{mathrsfs}
\usepackage{amssymb}
\usepackage{amsmath}
\usepackage{verbatim}
\usepackage{multirow}

\usepackage{subfigure}

\newcommand{\hzero}{\ensuremath{\PHiggslightzero}} 
\newcommand{\Hzero}{\ensuremath{\PHiggsheavyzero}} 
\newcommand{\CP}{\ensuremath{\mathcal{C}\mathcal{P}}}

\newcommand{\squark}{\ensuremath{\tilde{q}}}

\newcommand{\sw}{\ensuremath{s_W}}
\newcommand{\swd}{\ensuremath{s^2_W}}
\newcommand{\cw}{\ensuremath{c_W}}
\newcommand{\cwd}{\ensuremath{c^2_W}}

\newcommand{\mHHd}{\ensuremath{m^2_{\Hzero}}}

\newcommand{\mhd}{\ensuremath{m^2_{\hzero}}}

 \newcommand{\sad}{\ensuremath{\sin^2\alpha}}

 \newcommand{\self}{\ensuremath{\Sigma}}
 
 \newcommand{\drsing}{\ensuremath{\Delta r_{\text{sing}}}}
 \newcommand{\drsm}{\ensuremath{\Delta r_{\text{SM}}}}
  \newcommand{\drexp}{\ensuremath{\Delta r_{\text{exp}}}}
 \newcommand{\ddrsing}{\ensuremath{\delta(\Delta r_{\text{sing}})}}

 \newcommand{\deltarsing}{\ensuremath{\delta(\Delta r_{\text{sing}})}}
 \newcommand{\deltarhosing}{\ensuremath{\delta\rho_{\text{sing}}}}
 \newcommand{\ddeltarhosing}{\ensuremath{\Delta(\delta\rho_{\text{sing}})}}

\newcommand{\eqn}{equation}
\newcommand{\al}{\alpha}
\newcommand{\lb}{\left(}
\newcommand{\rb}{\right)}
\newcommand{\be}{\beta}

\newcommand{\higgssignals}{Bechtle:2013xfa,Stal:2013hwa,Bechtle:2014ewa}
\newcommand{\oblique}{Peskin:1990zt,Peskin:1991sw,Maksymyk:1993zm}
\begin{document}


\bibliographystyle{h-physrev}
\author{D. L\'opez-Val$^{a}$, T. Robens$^{b}$}
\title {$\Delta r$ and the W--boson mass in the 
    Singlet Extension of the Standard Model}
\maketitle

\begin{center}
\textit{$^{a}$ Center for Cosmology, Particle Physics and Phenomenology CP3 \\
Universit\'e Catholique de Louvain \\ 
Chemin du Cyclotron 2,  
B-1348 Louvain--la--Neuve,  Belgium}\\

\vskip5mm

\textit{$^{b}$ IKTP, Technische Universit\"at Dresden \\ 
Zellescher Weg 19, D-01069 Dresden, Germany}\\

\vskip5mm

E-mails: david.lopezval@uclouvain.be, tania.robens@tu-dresden.de
\vskip15mm

\end{center}

 \begin{quotation}
 The link between the electroweak gauge boson masses and the Fermi constant
 via the muon lifetime measurement is instrumental
 for constraining and eventually pinning down new physics. We consider the simplest extension of the Standard Model
 with an additional real scalar {$SU(2)_L\otimes\,U(1)_Y$} singlet
 and compute the electroweak precision parameter $\Delta r$,
 along with the corresponding theoretical prediction for the W--boson mass.
 When confronted with the experimental W--boson mass measurement, our predictions impose
 limits on the
 singlet model parameter space. We identify regions, especially in the mass range which is accessible by the LHC, where these correspond to the most stringent experimental constraints that are currently available.
 \end{quotation}
 
%

\section{Introduction}
The relation between the Electroweak (EW) gauge boson masses, the Fermi
constant [$G_F$] and the fine structure constant [$\alpha_{\text{em}}$]
is anchored experimentally via the muon lifetime measurement and constitutes a prominent
tool for testing the quantum structure of the Standard Model (SM) and its manifold conceivable
extensions. This relation is
conventionally expressed in the literature by means of the
$\Delta r$ parameter \cite{Kennedy:1988sn,Hollik:1988ii,Hollik:1993cg,Langacker:1996qb,Hollik:2003cj,Hollik:2006hd} and plays a major role in placing bounds on, and eventually unveiling 
new physics coupled to the standard electroweak Lagrangian.

\smallskip{}
Aside from being interesting on its own, the quantum effects traded by $\Delta r$ are part
of the electroweak radiative corrections to production and decay
processes in the SM and beyond. In particular, the knowledge of $\Delta r$ is
a required footstep towards a full one--loop electroweak characterization of the Higgs boson decay
modes in the singlet extension of the SM \cite{prep}.

\medskip{}
The calculation of {electroweak precision observables} (EWPO) and its role in constraining manifold extensions of the SM has been object of dedicated attention
in the literature 
\cite{Grifols:1983gu,Grifols:1984xs,Kennedy:1988sn,Hollik:1988ii,\oblique,Altarelli:1990zd,Altarelli:1991fk,Garcia:1994wu,Garcia:1994wv,Hollik:2003cj,Heinemeyer:2004gx,Hollik:2006hd,Wells:2005vk,Sirlin:2012mh}, 
{including in particular the singlet extension of the SM, cf. e.g. {Refs.~\cite{Bowen:2007ia,Profumo:2007wc,Barger:2007im,Dawson:2009yx,Cline:2009sn,Englert:2011yb,Gupta:2012mi,Bertolini:2012gu,Dolan:2012ac,Pruna:2013bma,Englert:2013tya,Chivukula:2013xka}} \footnote{{Cf. also \cite{Profumo:2014opa}, which appeared after the work presented here.}}}.
Theoretical predictions for $\Delta r$ and {for the W--boson mass $[m^{\text{th}}_W]$} were first
derived in the context of the SM ~\cite{Sirlin:1980nh,Marciano:1980pb} and 
later on extended
to new physics models {such as} the Two-Higgs--Doublet Model (2HDM) 
\cite{Frere:1982ma,Bertolini:1985ia,Hollik:1986gg,Hollik:1987fg,Froggatt:1991qw,He:2001tp,Grimus:2007if,Grimus:2008nb,LopezVal:2012zb} and the
Minimal Supersymmetric Standard Model (MSSM) 
\cite{vanderBij:1983bw,Barbieri:1989dc,Gosdzinsky:1990sk,Garcia:1993sb,Chankowski:1993eu,Freitas:2002ja,Heinemeyer:2002jq,Heinemeyer:2004gx,Heinemeyer:2006px,Heinemeyer:2013dia}. 
These predictions have proven to be relevant  
not only to {impose} parameter space constraints, 
but also to identify new physics structures
capable to in part reconcile the well--known tension
between the SM prediction 
and the experimental value, 
$|m_W^{\text{SM}}-m_W^{\text{exp}}| \simeq 20$ MeV.
For instance, in Ref.~\cite{LopezVal:2012zb} it was shown that the extended Higgs
sector of the general Two--Higgs--Doublet Model (2HDM) could yield
$m_W^{\text{2HDM}} \gtrsim m_W^{\text{SM}} $, thus potentially
alleviating the present discrepancy.

\smallskip{}
Our main 
endeavour {in this note} is to provide a one--loop
evaluation of the electroweak parameter $\Delta r$ and 
the W--boson mass in the presence
of {one extra} real {scalar} $SU(2)_L\otimes\,U(1)_Y$ singlet. This model, which incorporates an additional neutral, \CP-even spinless state, corresponds to 
the simplest 
renormalizable
extension of the SM, and can also be viewed as
an effective description of the low--energy Higgs sector of a more fundamental UV completion. 
Pioneered by Refs.~\cite{Silveira:1985rk,Schabinger:2005ei,Patt:2006fw}, this class of models has undergone
dedicated scrutiny for the past two
decades, revealing rich phenomenological implications, 
especially in the context of collider physics, ~see e.g. 
\cite{O'Connell:2006wi,BahatTreidel:2006kx,Barger:2007im, Bhattacharyya:2007pb, Gonderinger:2009jp, Dawson:2009yx, Bock:2010nz,Fox:2011qc, Englert:2011yb,Englert:2011us,Batell:2011pz, Englert:2011aa, Gupta:2011gd, Batell:2012mj, Dolan:2012ac, Bertolini:2012gu,Batell:2012mj,Bazzocchi:2012pp,Lopez-Val:2013yba,Heinemeyer:2013tqa,Chivukula:2013xka,Englert:2013tya,Cooper:2013kia,Caillol:2013gqa,Coimbra:2013qq,Eichhorn:2014qka}.   

\medskip{}

Our starting point is the current most precise theoretical prediction for the SM W--{boson} mass
[$m_W^{\text{SM}}$], which
is known exactly at two--loop accuracy, including up to leading
three--loop contributions 
\cite{Freitas:2002ja,Freitas:2002ve,Awramik:2002wn,Awramik:2002vu,Awramik:2003ee,Awramik:2003rn,Onishchenko:2002ve,vanderBij:2000cg}.
We combine these pure SM effects with the genuine singlet model {one--loop}
contributions {and analyse their dependences on the relevant model parameters}.
{Next we} correlate our results with the experimental measurement of the
W--boson {mass} and  
derive constraints
on the singlet model parameter space.
{Finally, we compare them} to {complementary constraints}
from direct collider searches,
as well as to 
the more conventional tests based on global fits
to electroweak precision observables.

\section{$\Delta r$ and $m_W$ as Electroweak precision measurements}
\label{sec:ewpo}

In the so--called ``$G_F$ scheme'', electroweak precision calculations
use the experimentally measured Z--boson mass [$m_Z$], the fine--structure constant at zero momentum [$\alpha_{\text{em}}(0)$],
and the Fermi constant [$G_F$] as input values. The latter is linked
to the muon lifetime via
\cite{Hollik:1993cg,Hollik:1988ii,Hollik:2003cj,Hollik:2006hd}
\begin{eqnarray}
\tau_{\mu}^{-1} = \frac{G_F^2 \, m_\mu^5}{192 \pi^3} \;
F\left(\frac{m_{\mathrm{e}}^2}{m_\mu^2}\right)
\left(1 + \frac{3}{5} \frac{m_\mu^2}{m_{\PW}^2} \right)
\left(1 + \Delta_{\rm QED} \right) ,
\label{eq:fermi}
\end{eqnarray}
\noindent where $F(x)=1-8x-12x^2\ln x+8x^3-x^4$. Following the standard
conventions in the literature, the above defining relation for $G_F$
includes  the finite QED
contributions $\Delta_{\rm QED}$ obtained within the Fermi Model -- which are known to
two--loop accuracy \cite{Behrends:1955mb,Kinoshita:1958ru,vanRitbergen:1999fi,Steinhauser:1999bx,Pak:2008qt}. 
Matching the muon {lifetime in the Fermi model} 
to the equivalent calculation within the full--fledged SM
yields the relation:

\begin{eqnarray}
m_{\PW}^2 \left(1 - \frac{m_{\PW}^2}{m_Z^2}\right)=
\frac{\pi \alpha_{\text{{em}}}}{\sqrt{2} G_F} \left(1 + \Delta r\right)\,
\quad \text{with} \quad \Delta r\equiv\frac{\hat\Sigma_{\PW}(0)}{m_W^2}+\Delta\,r^{[{\rm vert, box}]}\,,
\label{eq:deltar_def1}
\end{eqnarray}

\noindent which is
the conventional definition of $\Delta r$, with $m_{W,Z}$ being
the renormalized gauge boson masses in the on--shell scheme. 
Accordingly, we introduce the on--shell definition of the electroweak mixing angle  \cite{Sirlin:1980nh}
$\sin^2\theta_W\,=\,1-m_W^2/m_Z^2$, {along with the shorthand notations $\swd \equiv \sin^2\theta_W$, $c^2_W \equiv 1-\swd$.}
In turn, $\hat\Sigma_{\PW}(k^2)$ denotes the on--shell renormalized
$\PW$-boson self--energy. The latter accounts for the oblique part of the electroweak radiative corrections to the muon
decay. The non--universal (i.e. process--dependent) corrections rely
on the vertex and box contributions and
are subsumed into $\Delta\,r^{[{\rm vert, box}]}$. The
explicit expression for $\Delta r$ {after renormalization}
in the on--shell scheme may be written as a combination of loop
diagrams and counterterms as follows:
\begin{alignat}{5}
\Delta\,r &= \Pi_{\gamma}(0)- \frac{\cwd}{\swd}\,\left(\frac{\delta\,m_Z^2}{m_Z^2}
- \frac{\delta\,m^2_{\PW}}{m_{\PW}^2}\right) + \frac{\self_{\PW}(0)-\delta\,m_{\PW}^2}{m_{\PW}^2}
 + 2\,\frac{\cw}{\sw}\,\frac{\Sigma_{\gamma\PZ}(0)}{m_Z^2}+\Delta\,r^{[{\rm vert, box}]}
\label{eq:deltar_def3},
\end{alignat}
where {$\Pi_{\gamma}(0)$ stands for
the photon vacuum polarization, while} $\delta m^2_{W,Z}$ denote the {gauge boson} 
mass counterterms. Additional degrees of freedom and/or modified interactions will 
enter the loop diagrams describing the muon decay, making
$\Delta r$ (and so $m_W$) model--dependent quantities.
At present, the calculation of $\Delta r$ in the SM is complete
up to two loops
\cite{Djouadi:1987gn,Djouadi:1987di,Halzen:1990je,Halzen:1991ik,Kniehl:1991gu,Kniehl:1992dx,Djouadi:1993ss,Freitas:2000gg,Freitas:2002ja,Freitas:2002ve,Awramik:2002wn,Onishchenko:2002ve,Awramik:2002vu,Awramik:2002wv,Awramik:2003ee,Awramik:2003rn}
and includes also the leading three \cite{Avdeev:1994db,Chetyrkin:1995ix,Chetyrkin:1995js,Chetyrkin:1996cf,vanderBij:2000cg,Boughezal:2004ef} and four--loop pieces \cite{Boughezal:2006xk,Chetyrkin:2006bj}. 
The {dominant contribution stems from QED fermion loop corrections}
and is absorbed into the renormalization group running of the fine structure constant.

\bigskip{}
Taking $m_Z$ and $G_F$ as experimental
inputs, and using Eq.~\eqref{eq:deltar_def1}, the evaluation
of $\Delta r$ within the SM or beyond can
be translated into a {theoretical} prediction for the W--boson
mass $[m^{\rm
th}_{\PW}]$. For this we
need to (iteratively) solve the equation
\begin{eqnarray}
m_{\PW}^2 = \frac12\,m_Z^2\,\left[1 + \sqrt{1 - \frac{4\,\pi\alpha_{{\text{em}}}}{\sqrt{2}\,G_F\,m_Z^2}
\,[1 + \Delta\,r(m_{\PW}^2)]} \right]
\label{eq:mwpred}.
\end{eqnarray}
To first--order accuracy,
Eq.\,\eqref{eq:mwpred} implies that a shift $\delta(\Delta r)$ 
promotes to the $\PW$--boson mass through
\begin{equation}\label{eq:mwshift}
{\Delta  m_{\PW}}\simeq -\frac{1}{2}\,m_{\PW}\,\frac{\swd}{\cwd-\swd}\,\delta(\Delta r)\,.
\end{equation}
For $\Delta r=0$ one retrieves the tree-level value $m_W^{\rm tree}\simeq
80.94$ GeV. But the full theoretical result is smaller in the SM
since quantum effects yield {$\Delta r > 0$} of order few percent. 
Once we identify
the $\sim 126$ GeV resonance with the SM Higgs boson, all experimental
input values in Eq.~\eqref{eq:mwpred} are fixed 
and thereby the theoretical prediction for the W--boson mass is fully determined. {Setting} the SM Higgs boson mass to the 
{\sc HiggsSignals}\cite{\higgssignals} best--fit value $m_{\PHiggs} = 125.7$ GeV \cite{Bechtle:2013xfa}, 
one gets $\Delta r\simeq{0.038}$ $>0$, wherefrom $m_W^{\text{SM}} = 80.360 \, \GeV$. 
The estimated
theoretical uncertainty reads $\Delta m^{\text{th}}_W \simeq 4\,$ MeV \cite{Awramik:2003rn}
and stems mainly from the top mass measurement ~\cite{Heinemeyer:2003ud}.
This prediction needs to be confronted with the experimental
W--boson mass
measurement, whose present world--average combines 
the available results from LEP \cite{Alcaraz:2006mx}, CDF \cite{Aaltonen:2012bp} and D0 \cite{D0:2013jba} and renders

\begin{alignat}{5}
 m_W^{\text{exp}} = 80.385 \pm 0.015 \,\GeV \label{eq:mwexp}.
\end{alignat}

\noindent This represents an accuracy {at the $\simeq 0.02 \%$ level}.
The corresponding discrepancy with the {SM} theoretical prediction
$|m_W^{\text{exp}} - m_W^{\text{SM}}| \simeq$ 20 MeV
falls within the $1\sigma$--level ballpark; however, it is as large
as roughly 5 times the estimated theoretical error.
On the other hand, {these differences} {should be accessible by} the upcoming $\PW$--boson
mass measurements at the LHC, which are expected to pull the current uncertainty
down to $\Delta m^{\rm
exp}_{\PW} \simeq 10\,\MeV$  \cite{Bozzi:2011ww,Bernaciak:2012hj}. Furthermore, a high--luminosity
linear collider running in a low--energy mode at the $\PW^+\PW^-$ threshold
should be able to reduce it even further, namely at the level of
{$\Delta m_{\PW}^{\rm exp} \simeq 5$~MeV or even below \cite{Baak:2013fwa}}. 
This strongly justifies,
if not simply demands, precision calculations of $\Delta r$ and $m_W$ 
to probe, constrain, or even unveil, new physics structures linked
to the electroweak sector of the SM.

\bigskip{}
As a byproduct, the task of computing $\Delta r$ involves the evaluation of the so--called $\delta\rho$ parameter
\cite{Ross:1975fq,Veltman:1976rt,Veltman:1977kh,Einhorn:1981cy}.
The latter is defined upon the static contribution to the gauge boson
self--energies,

\begin{equation}
\frac{\Sigma_{\PZ}(0)}{m_{\PZ}^2} -
\frac{\Sigma_{\PW}(0)}{m_{\PW}^2}\equiv\delta\rho
\label{eq:deltarho-def1},
\end{equation}

\noindent and measures the ratio of the
neutral--to--charged weak current strength. 
Quantum effects yielding $\delta\rho \neq 0$ may be traced back to
the mass splitting between the partners of a given weak
isospin doublet, and so to the degree of departure 
from the global custodial $SU(2)$ invariance
of the SM Lagrangian. 
The $\delta\rho$ parameter is finite for each doublet of SM matter fermions and
is dominated by the top quark loops 

\begin{alignat}{5}
 \delta\rho_{\text{{SM}}}^{[t]} = \cfrac{3G_F m_t^2}{8\sqrt{2}\,\pi^2} \label{eq:deltathotop}.
\end{alignat}

\noindent In terms of $\delta\rho$, the general expression for
$\Delta r$ {can be recast as} \cite{Hollik:1988ii,Hollik:2003cj,Hollik:2006hd}:
\begin{equation}
\Delta r=\Delta\alpha-\frac{\cwd}{\swd}\,\delta\rho+\Delta r_{\rm rem} = \Delta\alpha + \Delta r^{[\delta\rho]}+\Delta r_{\rm rem}\,,
\label{eq:deltar_def4}
\end{equation}

\noindent where $\Delta r^{[\delta\rho]} \equiv
-(\cwd/\swd)\delta\rho$ denotes the individual contribution 
from the static part of the self--energies. 
The $\Delta \alpha$ piece accounts for the (leading) QED light--fermion corrections,
while the so--called ``remainder'' term [$\Delta r_{\text{rem}}$] condenses
the remaining (though not negligible) effects. In fact, {in the SM} we have
$\Delta \alpha \simeq 0.06$ and $\Delta r_{\text{rem}} \simeq 0.01$,
while $\Delta r^{[\delta\rho]} \simeq -0.03$ \cite{Hollik:1993cg,Hollik:2003cj,Hollik:2006hd}.

\medskip{}

\noindent At variance with this significant contribution,
the counterpart
Higgs boson--mediated effects are comparably milder in the SM
and feature a trademark logarithmic dependence on the Higgs mass \cite{Veltman:1976rt}
\footnote{One should bear in mind that 
the Higgs boson contribution in the SM [$\delta\rho^{[\PHiggs]}_{\text{SM}}$] is 
neither UV finite nor gauge invariant on its own, but only in combination with the
remaining bosonic contributions.},


\begin{alignat}{5}
\delta\rho^{[\PHiggs]}\simeq -\frac{3\sqrt{2}\,G_F\,m_{\PW}^2}{16\,\pi^2}\,\frac{\swd}{\cwd}\,\left\{\ln\frac{m_H^2}{m_W^2}-\frac56\right\}+...
\label{eq:deltarhohiggs}\quad .
\end{alignat}

\noindent {Remarkably}, this telltale \emph{screening} behavior
 does not hold in general for extended Higgs sectors -- viz. in the general 2HDM \cite{LopezVal:2012zb}.

\section{$\Delta r$ and $m_W$ in the singlet model}
\label{sec:ewposinglet}

\subsection{Model parametrization at leading--order}
\label{sec:model}

Our starting point is the most general form of the 
gauge invariant, renormalizable potential
involving one real {$SU(2)_L\otimes\,U(1)_Y$} singlet $S$ and one
doublet $\Phi$, the latter carrying the quantum numbers of the SM Higgs {weak isospin} doublet (see e.g. \cite{Schabinger:2005ei,Patt:2006fw,Pruna:2013bma}):
\begin{equation}\label{lag:s}
\mathscr{L}_s = \left( \mathcal{D}^{\mu} \Phi \right) ^{\dagger} \mathcal{D}_{\mu} \Phi + 
\partial^{\mu} S \partial_{\mu} S - V(\Phi,S ) \, ,
\end{equation}
with the potential
\begin{eqnarray}\label{eq:potential}\nonumber
V(\Phi,S ) &=& -\mu_1^2 \Phi^{\dagger} \Phi - \mu_2^2  S ^2 +
\left(
\begin{array}{cc}
\Phi^{\dagger} \Phi &  S ^2
\end{array}
\right)
\left(
\begin{array}{cc}
\lambda_1 & \frac{\lambda_3}{2} \\
\frac{\lambda_3}{2} & \lambda _2 \\
\end{array}
\right)
\left(
\begin{array}{c}
\Phi^{\dagger} \Phi \\  S^2 \\
\end{array}
\right) \\
\nonumber \\ 
&=& -\mu_1^2 \Phi^{\dagger} \Phi -\mu_2^2 S ^2 + \lambda_1
(\Phi^{\dagger} \Phi)^2 + \lambda_2  S^4 + \lambda_3 \Phi^{\dagger}
\Phi S ^2.
\end{eqnarray}

\noindent 
{For the sake of simplicity we consider a minimal version of the singlet model, 
with an additional $\mathcal{Z}_2$ symmetry 
forbidding additional terms in the potential. 
We allow both of the scalar fields to acquire a Vacuum Expectation Value (VEV), in which
case the $\mathcal{Z}_2$ symmetry is spontaneously broken by the singlet VEV.  
The breaking of such a discrete symmetry 
during the electroweak phase transition in the early universe may in principle lead
to problematic weak--scale cosmic domain walls ~\cite{Kobzarev:1974cp,Kibble:1976sj,Kibble:1980mv}. However, analyses of the stability
and evolution of such topological defects in multiscalar extensions
of the SM (cf. e.g. Refs.~\cite{Preskill:1991kd,Abel:1995wk,Panagiotakopoulos:1998yw}) 
identify a variety of mechanisms that may sidestep these
issues. These can also be evaded by extending this minimal setup with additional $\mathcal{Z}_2$ breaking terms~\cite{Barger:2008jx},
which would nevertheless have no direct impact on our analysis. In this sense, let us emphasize that we interpret  
the singlet model as the low--energy effective Higgs sector of a more fundamental UV--completion (cf. e.g. 
a model with an extended gauge group \cite{Basso:2010yz,Basso:2011na}), whose specific details are either way not relevant for the purposes of our study.}

\medskip{}
The neutral components of these fields can be expanded around their
respective VEVs as follows:

\begin{equation}
 \Phi = \left(\begin{array}{c}  
  G^{\pm} \\ \cfrac{v_d + l^0 + iG^0}{\sqrt{2}}
 \end{array}\right) \qquad \qquad S = \cfrac{v_s + s^0}{\sqrt{2}}
 \label{eq:components}.
\end{equation}

\noindent The minimum of the above potential is achieved under the conditions

\begin{alignat}{5}
 \mu^2_1 = \lambda_1 v_d^2 + \cfrac{\lambda_3 v_s^2}{2}; \qquad \qquad 
  \mu^2_2 = \lambda_2 v_s^2 + \cfrac{\lambda_3 v^2_d}{2}
  \label{eq:minimum},
\end{alignat}

\noindent while the quadratic terms in the fields generate the mass--squared matrix

\begin{equation}
\mathcal{M}^2_{{ls}} =  \left( \begin{array}{cc}  
      2\lambda_1\,v_d^2 & \lambda_3\,v_d\,v_s \\
      \lambda_3\,v_d\,v_s & 2\lambda_2\,v_s^2  
     \end{array} \right)
\label{eq:mass-matrix}.
\end{equation}

\noindent Requiring this matrix to be positively--defined leads
to the stability conditions\footnote{Cf. e.g. \cite{Pruna:2013bma} for a more detailed discussion.}

\begin{alignat}{5}
 \lambda_1, \lambda_2 > 0; \qquad 4\lambda_1\lambda_2 - \lambda_3^2 > 0 \label{eq:stability}\; .
\end{alignat}

The above mass matrix in the gauge basis $\mathcal{M}^2_{ls}$
can be transformed into the (tree--level) mass basis through the rotation
$R(\alpha)\,\mathcal{M}^2_{ls}\,R^{-1}(\alpha) = \mathcal{M}^2_{hH} = \text{diag}(m_{\hzero}^2 \; m^2_{\Hzero})$,
with

\begin{equation}
R(\alpha) = \left( \begin{array}{cc} \cos\alpha & -\sin\alpha \\ \sin\alpha & \cos\alpha 
 \end{array} \right) \qquad \text{and} \qquad 
 \tan(2\alpha) = \cfrac{\lambda_3v_dv_s}{\lambda_1 v_d^2 - \lambda_2v_s^2}
 \label{eq:rotation}\, .
\end{equation}

\noindent Its eigenvalues then read

\begin{alignat}{5}
 m^2_{\hzero,\Hzero} &= \lambda_1\,v_d^2 + 
 \lambda_2\,v_s^2 \mp |\lambda_1\,v_d^2 - 
 \lambda_2\,v_s^2|\,\sqrt{1+\tan^2(2\alpha)}
 \quad \text{with the convention} \quad m_{\Hzero}^2 > m_{\hzero}^2 \label{eq:masseigen},
\end{alignat}

\noindent and correspond to a 
light [$\hzero$] and a heavy [$\Hzero$] $\mathcal{CP}$-even  mass--eigenstate. From Eq.~\eqref{eq:rotation}, we see that
both are admixtures of the doublet [$l^0$] and the singlet [$s^0$] neutral
components  

\begin{alignat}{5}
\hzero =l^0  \cos\alpha  -s^0 \sin\alpha  \qquad \text{and} \qquad
 \Hzero &= l^0 \sin\alpha  +s^0 \cos\alpha . \label{eq:masseigen-repeat}
\end{alignat}

{The Higgs sector in this model} is determined by five independent parameters,  which can be chosen as
\begin{\eqn*}
m_{\hzero},\,m_{\Hzero}, \,\sin\al, \,{v_d}, \,\tan\be\,\equiv\,\frac{v_d}{v_s}\, ,
\end{\eqn*}
where the doublet VEV is fixed in terms of the Fermi constant through $v_d^2 = G_F^{-1}/\sqrt{2}$. 
Furthermore, we fix one of the Higgs masses to the LHC value of $125.7\,\GeV$; therefore, three parameters of the model are presently not determined by any experimental measurement. \\

As only the doublet component
can couple to the fermions (via ordinary Yukawa interactions) and the gauge bosons
(via the gauge covariant derivative), all of the Higgs couplings to SM particles
are rescaled universally, {yielding}

\begin{alignat}{5}
 g_{xxh} = g_{xxh}^{\text{SM}}(1 + \Delta_{xh}) \qquad \text{with} \qquad 1+\Delta_{xh} = 
 \begin{cases}\cos\alpha& h\,=\,\hzero\\ \sin\al&h=\Hzero \end{cases} \label{eq:coupling}. 
\end{alignat}

\subsection{Calculation details}
\label{sec:details}


{Let us now focus on} the calculation of $\Delta r$ and $m_W$ in the singlet extension of the SM.
The pure SM contributions [$\Delta r_{\text{SM}}$]
and the genuine singlet model effects [$\delta(\drsing)$] can be split into  
two UV-finite, gauge--invariant subsets  and treated separately:

\begin{alignat}{5}
 \drsing &= \drsm + \deltarsing \label{eq:drsing}. 
\end{alignat}

We here include
the state--of--the--art $\Delta r_{\text{SM}}$ evaluation, 
extracted from Eq.~\eqref{eq:deltar_def1} and
the numerical parametrization given in
Ref.\cite{Awramik:2003rn}, which renders the central values

\begin{\eqn}\label{eq:sm_val}
m_W^\text{SM}\,=\,80.360\,\GeV \qquad \text{and} \qquad 
\drsm\,=\,37.939\times 10^{-3}.
\end{\eqn}

We set the top-quark mass [$m_t = 173.07$ GeV] and the Z-boson mass  [$m_{\PZ} = 91.1875$ GeV] 
at their current best average values \cite{Beringer:1900zz}. 
The SM Higgs mass is fixed to the {\sc HiggsSignals} best--fit
value {of 125.7 GeV}.
This result for $\Delta r_{\text{SM}}$ includes 
the full set of available contributions, {combining} the full--fledged
two--loop bosonic \cite{Onishchenko:2002ve,Awramik:2002wv} and 
fermionic \cite{Freitas:2000gg,Freitas:2002ja,Awramik:2003rn} effects,
alonside the leading three--loop corrections at $\mathcal{O}(G_F^3 m^6_t)$
and $\mathcal{O}(G_F^2\,\alpha_s\,m^4_t)$ \cite{vanderBij:2000cg}. 

\begin{figure}[t!]
\begin{center}
\includegraphics[width=1\textwidth]{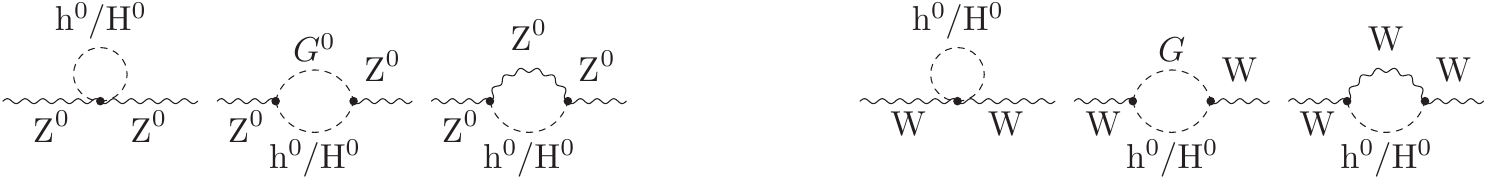}
\caption{\label{fig:oneloop} One--loop Higgs boson--mediated contributions to the 
weak gauge boson self--energies in the 
singlet model.  The charged
and neutral Goldstone boson contributions appear explicitly 
in the 'tHooft-Feynman gauge. The Feynman diagrams are generated using {\sc FeynArts.sty} \cite{Hahn:2000kx}.}
\end{center}
\end{figure}

\smallskip{} 
The genuine singlet model contributions [$\delta(\Delta r_{\text{sing}})$]
originate from the Higgs--boson mediated loops
building up the weak gauge boson self--energies,
{which are shown in Fig.~\ref{fig:oneloop}.}
This model--dependent part relies on the Higgs
masses [$m_{\hzero}, m_{\Hzero}$] and the mixing angle [$\sin\alpha$], and {we compute it} analytically to one--loop order. {As the Higgs self--interactions
do not feature at one--loop, the results are insensitive to $\tan\beta$.}

\smallskip{}
At this point, care must be taken not to double--count the {pure SM}
Higgs--mediated contributions. 
To that aim we define [$\ddrsing$] in Eq. (\ref{eq:drsing})
 upon subtraction of the SM contribution:

\begin{alignat}{5}
 \delta(\drsing) \equiv \drsing^{[\PHiggs]} - \drsm^{[\PHiggs]} \quad 
 \text{where} \quad  \drsm^{[\PHiggs]} = \drsing^{[\PHiggs]}\Big{\lvert}_{\sin\alpha=0}
 \label{eq:subtractoverlap},
\end{alignat}

\noindent {while} the superscript $[\PHiggs]$ selects the Higgs--mediated
contributions in each case. 
In {this expression we explicitly identify}
the {SM--like Higgs boson} with 
the lighter of the two mass--eigenstates $[\hzero]$,  
while the second eigenstate $[\Hzero]$ is assumed to describe a (so far unobserved) heavier Higgs companion. 
Analogous expressions can be derived for the complementary case [$m_{\Hzero}\,=\,125.7\,\GeV > m_{\hzero}$],
{wherein the SM limit corresponds to $\cos\alpha = 0$}.
The phenomenology of both {possibilities} 
is analysed separately in section \ref{sec:numerical}.

\medskip{} {With this in mind,} the {purely}
singlet model contributions to the gauge boson self--energies give

\begin{alignat}{5}
 \overline{\self}_{\PZ\PZ}(p^2) &= \cfrac{\alpha_{\text{em}}\,\sad}{4\,\pi\swd\cwd}\,
\Bigg\{\cfrac{\left[A_0(\mHHd) - A_0(\mhd)\right]}{4}\,
+m_Z^2\,\left[B_0(p^2,\mHHd,m_Z^2)-B_0(p^2,\mhd,m_Z^2) \right] \notag \\
& -\left[B_{00}(p^2,\mHHd,m_Z^2)-B_{00}(p^2,\mhd,m_Z^2) \right] \Bigg{\}} 
\label{eq:selfzz}
\end{alignat}

\begin{alignat}{5}
 \overline{\self}_{\PW\PW}(p^2) &= \cfrac{\alpha_{\text{em}}\,\sad}{4\,\pi\swd}\,
\Bigg\{\cfrac{\left[A_0(\mHHd) - A_0(\mhd)\right]}{4}\,
+m_W^2\,\left[B_0(p^2,\mHHd,m_W^2)-B_0(p^2,\mhd,m_W^2) \right] \notag \\
& -\left[B_{00}(p^2,\mHHd,m_W^2)-B_{00}(p^2,\mhd,m_W^2) \right] \Bigg{\}} 
\label{eq:selfww}.
\end{alignat}

\noindent The loop integrals in the above equations 
are expressed in terms of the standard Passarino--Veltman
coefficients in the conventions of \cite{Hahn:1998yk}.
The overlined notation $\overline{\Sigma}$ indicates
that the overlap with the SM Higgs--mediated contribution 
has been removed according to Eq.~\eqref{eq:subtractoverlap}. 
Analogous
expressions where
$[m_{\hzero} \leftrightarrow m_{\Hzero}]$ and $[\cos\alpha \leftrightarrow \sin\alpha]$
are valid if we identify the heavy
scalar eigenstate [$\Hzero$] with the SM--like
Higgs boson. 

\medskip{}
{The presence of the additional singlet has a twofold
impact:}
i) first, via the novel {one--loop diagrams mediated by the exchange of the additional Higgs boson,}
as displayed in Fig.~\ref{fig:oneloop};
ii) second, via the {reduced} coupling strength of the SM--like Higgs
to the weak gauge bosons, rescaled by the mixing angle (cf. Eq.~\ref{eq:coupling}). 

\medskip{}
At this stage, 
we in fact do not yet have to specify a 
complete renormalization scheme for the model. It suffices to consider
the weak gauge boson field and mass renormalization entering Eq.~\eqref{eq:deltar_def1}.
The relevant counterterms therewith are fixed in the on--shell scheme 
\cite{Sirlin:1980nh,Bohm:1986rj,Hollik:1988ii,Denner:1991kt}, 
i.e. by requiring the real part of the transverse renormalized self--energies to vanish at the respective
gauge boson pole masses, while setting the propagator residues to unity:
\begin{eqnarray*}
\text{Re}\hat{\Sigma}^W_T(m_W^2)\,=\,0, &&\text{Re}\hat{\Sigma}^Z_T(m_Z^2)\,=\,0,\\
\text{Re}\frac{\partial\hat{\Sigma}^W_T(p^2)}{\partial p^2}\Big{|}_{p^2=m_W^2}\,=\,0, & &\text{Re}\frac{\partial\hat{\Sigma}^Z_T(p^2)}{\partial p^2}\Big{|}_{p^2=m_Z^2}\,=\,0\; .
\end{eqnarray*}

\medskip{}
The use of the on--shell scheme, which is customary in this context, provides 
an unambiguous meaning to the free parameters of the model, allowing for
a direct mapping between the bare parameters in the classical Lagrangian
and the physically measurable quantities in the quantized renormalizable Lagrangian.
For instance, choosing on--shell renormalization conditions ensures that the weak
gauge boson masses in Eqs.~\eqref{eq:fermi}-\eqref{eq:deltar_def1} correspond to their physical masses \footnote{On--shell mass renormalization
in theories with mixing between the gauge eigenstates, 
as in the Higgs sector of the singlet model, must be {nonetheless} addressed with care.
In these cases, 
quantum effects generate off--diagonal terms in the loop--corrected
propagators, which can be absorbed into the renormalization of the mixing angle.
However, it can be shown that, regardless of the specific renormalization
scheme chosen for the mixing angle, the on--shell renormalized masses coincide
with the physical (pole) masses to one--loop accuracy. A detailed discussion
on this issue as well as on the complete renormalization scheme as such for the singlet model 
will be presented in \cite{prep}. 
}.

\medskip{}
The complete singlet model prediction in Eq.~\eqref{eq:drsing} is exact 
to one--loop order and, as alluded to above, it includes in addition all known
higher order SM effects up to leading three--loop precision.
Finally, let us also remark that the additional singlet--mediated
contributions to the vertex and box diagrams contained in $\Delta r^{[\text{vert,box}]}$ (cf. Eq.~\eqref{eq:deltar_def3})
are {suppressed by the light fermion Yukawa couplings and therefore negligible.

\medskip
In turn, the static contributions traded by the $\delta\rho$ parameter, as defined in Eq.~(\ref{eq:deltarho-def1}),
can be obtained by taking the limit $p^2 \to 0$ on Eqs.~\eqref{eq:selfzz}-\eqref{eq:selfww} and are given by

\begin{alignat}{5}
 \ddeltarhosing &\equiv \deltarhosing^{[\PHiggs]}-\delta\rho^{[\PHiggs]}_{\text{SM}} \notag \\
& \cfrac{G_F\,\sad}{2\sqrt{2}\pi^2}\,\Bigg{\{}
m_Z^2\,\left[ 
\log\left(\cfrac{\mhd}{\mHHd}\right) + \cfrac{m_Z^2}{\mhd-m_Z^2}\,
\log\left(\cfrac{\mhd}{m_Z^2}\right) -
\cfrac{m_Z^2}{\mHHd-m_Z^2}\,
\log\left(\cfrac{\mHHd}{m_Z^2}\right) \right. \notag \\
& \left. \qquad + \cfrac{\mHHd}{4(\mHHd-m_Z^2)}\,
\log\left(\cfrac{\mHHd}{m_Z^2}\right) 
- \cfrac{\mhd}{4(\mhd-m_Z^2)}\,
\log\left(\cfrac{\mhd}{m_Z^2}\right)
\right] \notag \\
& \qquad - m_W^2\,\left[ 
\log\left(\cfrac{\mhd}{\mHHd}\right) + \cfrac{m_W^2}{\mhd-m_W^2}\,
\log\left(\cfrac{\mhd}{m_W^2}\right) -
\cfrac{m_W^2}{\mHHd-m_W^2}\,
\log\left(\cfrac{\mHHd}{m_W^2}\right) \right. \notag \\
& \left. \qquad + \cfrac{\mHHd}{4(\mHHd-m_W^2)}\,
\log\left(\cfrac{\mHHd}{m_W^2}\right) 
- \cfrac{\mhd}{4(\mhd-m_W^2)}\,
\log\left(\cfrac{\mhd}{m_W^2}\right)
\right]
\Bigg{\}}
\label{eq:deltarho-singlet2}
\end{alignat}

\noindent 
{(cf. also the expression for the $T$-parameter in the $\overline{MS}$ scheme \cite{Profumo:2007wc}
\footnote{{It is easy to check that Eq.~\eqref{eq:deltarho-singlet2} is equivalent
to Eq.~(5.1) of Ref.~\cite{Profumo:2007wc}, recalling
that in our case we identify $m_{\hzero}$ with the SM Higgs mass.}}).
The logarithmic dependence
on both the light and the heavy Higgs masses follows the same screening--like 
pattern of the SM, as shown in Eq.~\eqref{eq:deltarhohiggs}. 
The {model--specific new physics imprints} are again to be found in
i) the additional Higgs contribution;
and ii) the {universally rescaled} Higgs couplings to the gauge bosons.
The size
of $\Delta(\delta\rho_{\text{sing}})$ is controlled by the overall factor
$\sim \sin^2\alpha$,  
while its sign, 
which is fixed by the respective Higgs and gauge boson mass {ratios}, is negative 
in all cases. {Equation~\eqref{eq:deltarho-singlet2}} 
therefore predicts a systematic,
negative yield from the {new physics effects} [$\Delta(\delta\rho_{\text{sing}}) < 0$],
which implies $\delta\rho_{\text{sing}} <  \delta\rho_{\text{SM}}$.
Finally, and owing to the fact that $\delta\rho$ is linked to $\Delta r$ via
Eq.~\eqref{eq:deltar_def4}, we may foresee  
$\drsing \equiv \drsm + \deltarsing > \drsm$
and hence $m_W^{\text{sing}} < m_W^{\text{SM}}$.
Keeping in mind the current $|m_W^{\text{exp}} - m_W^{\text{SM}}| \simeq 20$ MeV ($1\sigma$ level)
tension, this result anticipates tight constraints on
the singlet model parameter space -- at the level of, if not 
stronger than, those stemming from the {global fits
based on the oblique parameters $[S,T,U]$} \cite{\oblique} (cf. discussion in section \ref{sec:comparison}).
Conversely, when considering $m_{\Hzero}\,\sim\,126\,\GeV$ {and a light Higgs companion}  [$\hzero$], 
{similar arguments} {predict}
a systematic upward shift [$\Delta(\delta\rho_{\text{sing}}) > 0$] with a global $\cos^2\alpha$
rescaling. In this case, the {singlet model} has the potential to bring 
the theoretical value {[$m_W^{\text{sing}}$]} closer to the experimental measurement [$m^{\text{exp}}_W$].
In the next subsection we quantitatively justify all these statements.
\subsection{Numerical analysis}
\label{sec:numerical}

\begin{figure}[t!]
\begin{center}
\begin{tabular}{ccc}
%
\includegraphics[width=0.3\textwidth]{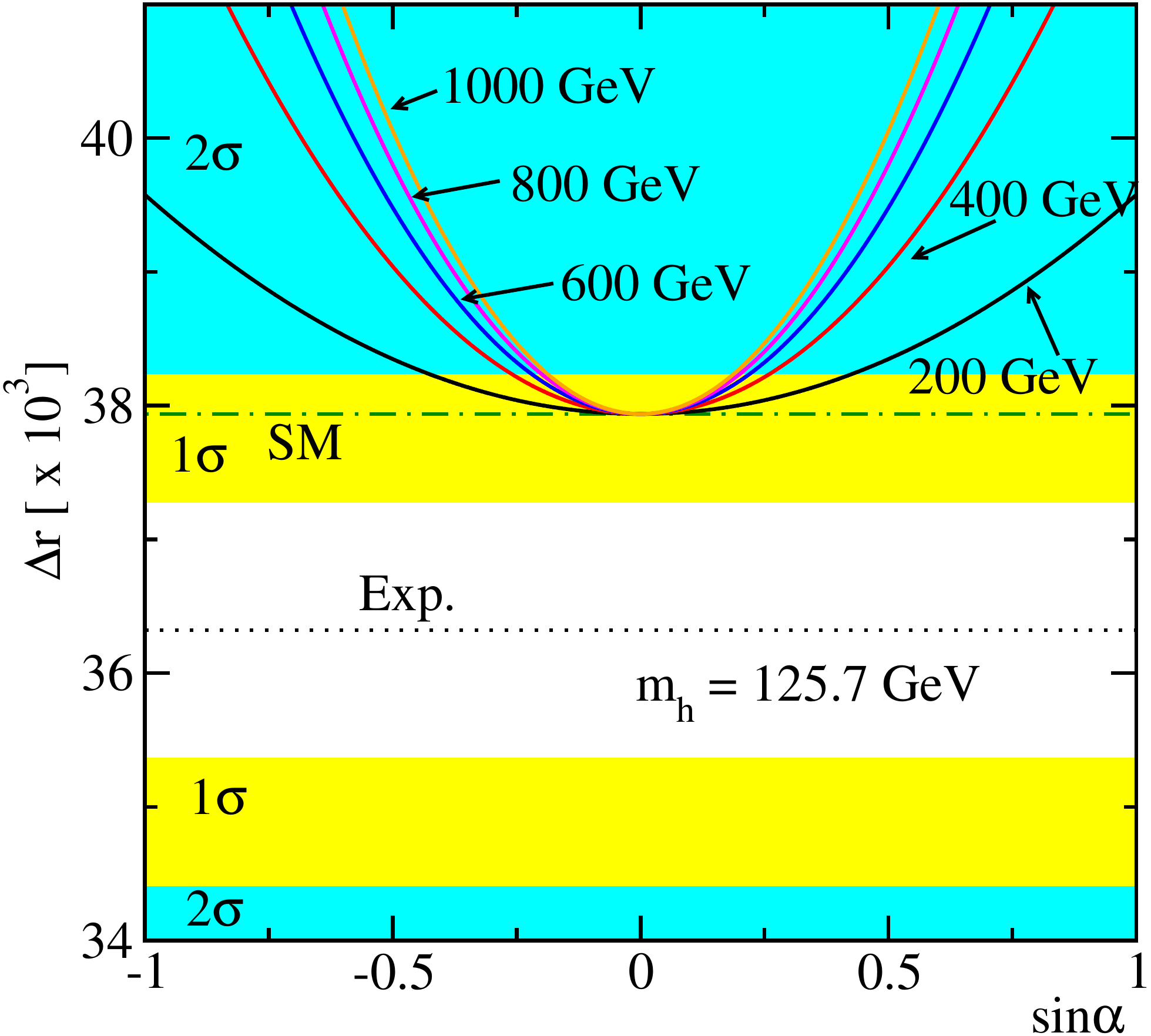} & & 
\includegraphics[width=0.31\textwidth]{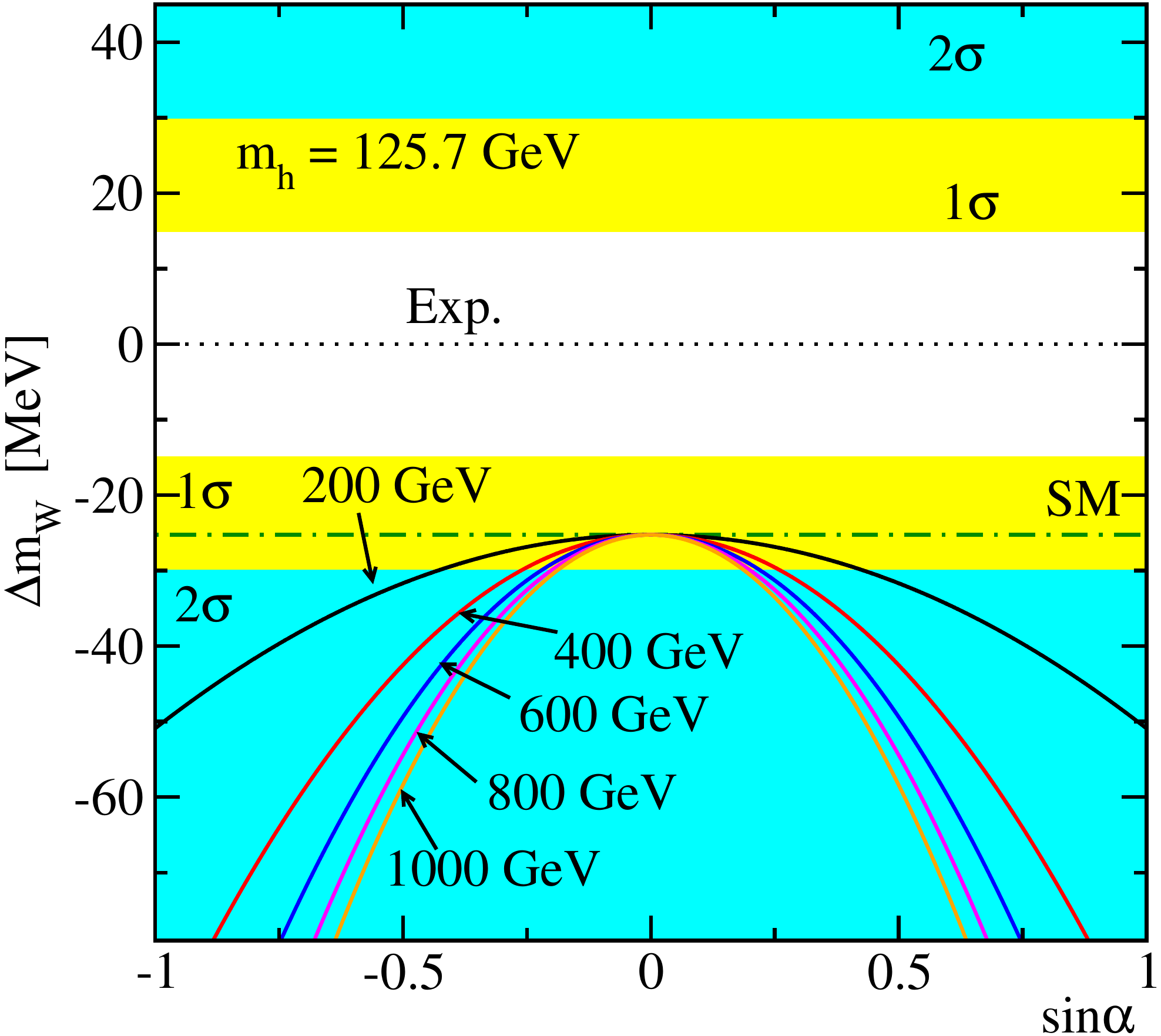} \\ & & \\
\includegraphics[width=0.3\textwidth]{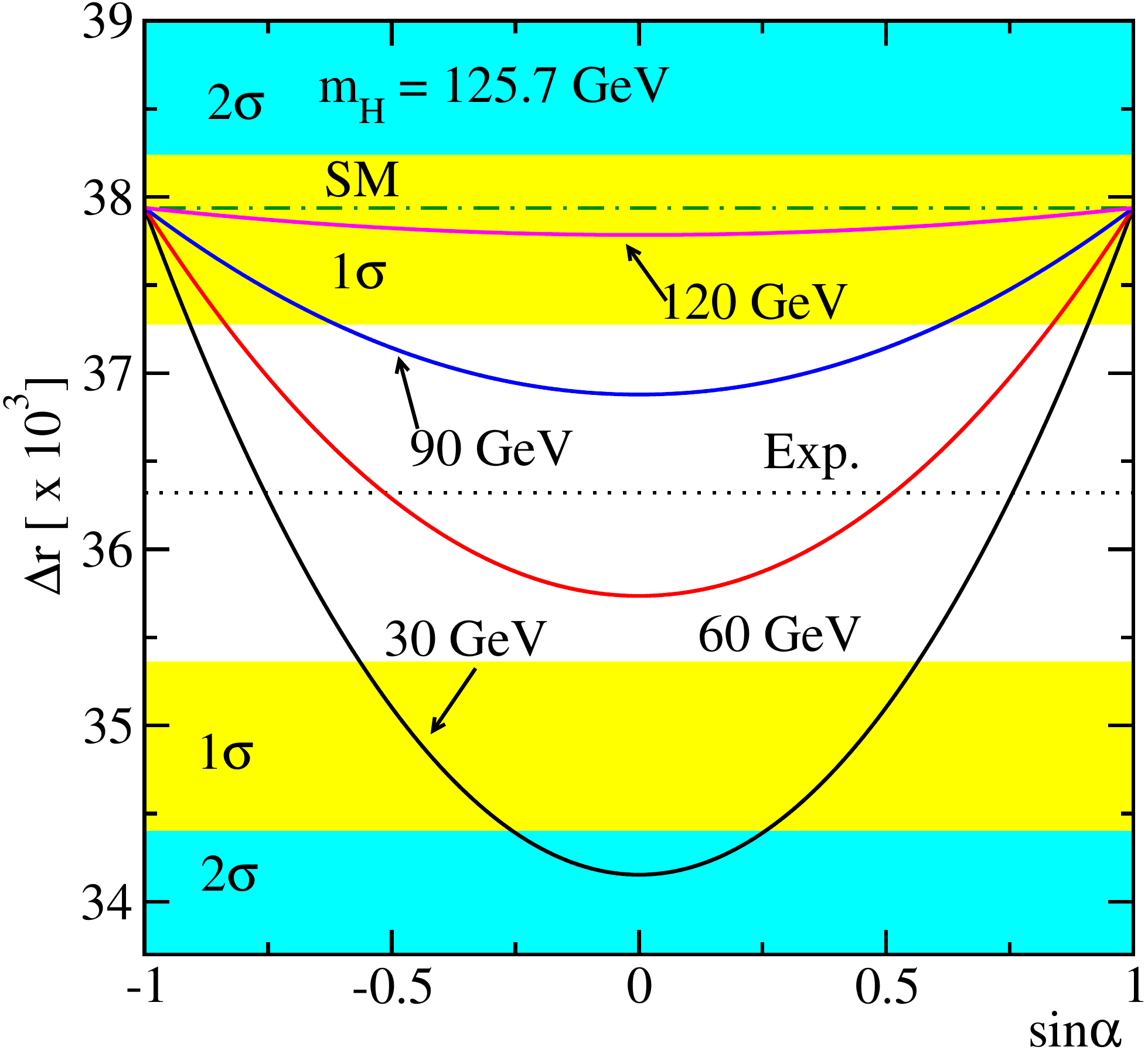} & & 
\includegraphics[width=0.31\textwidth]{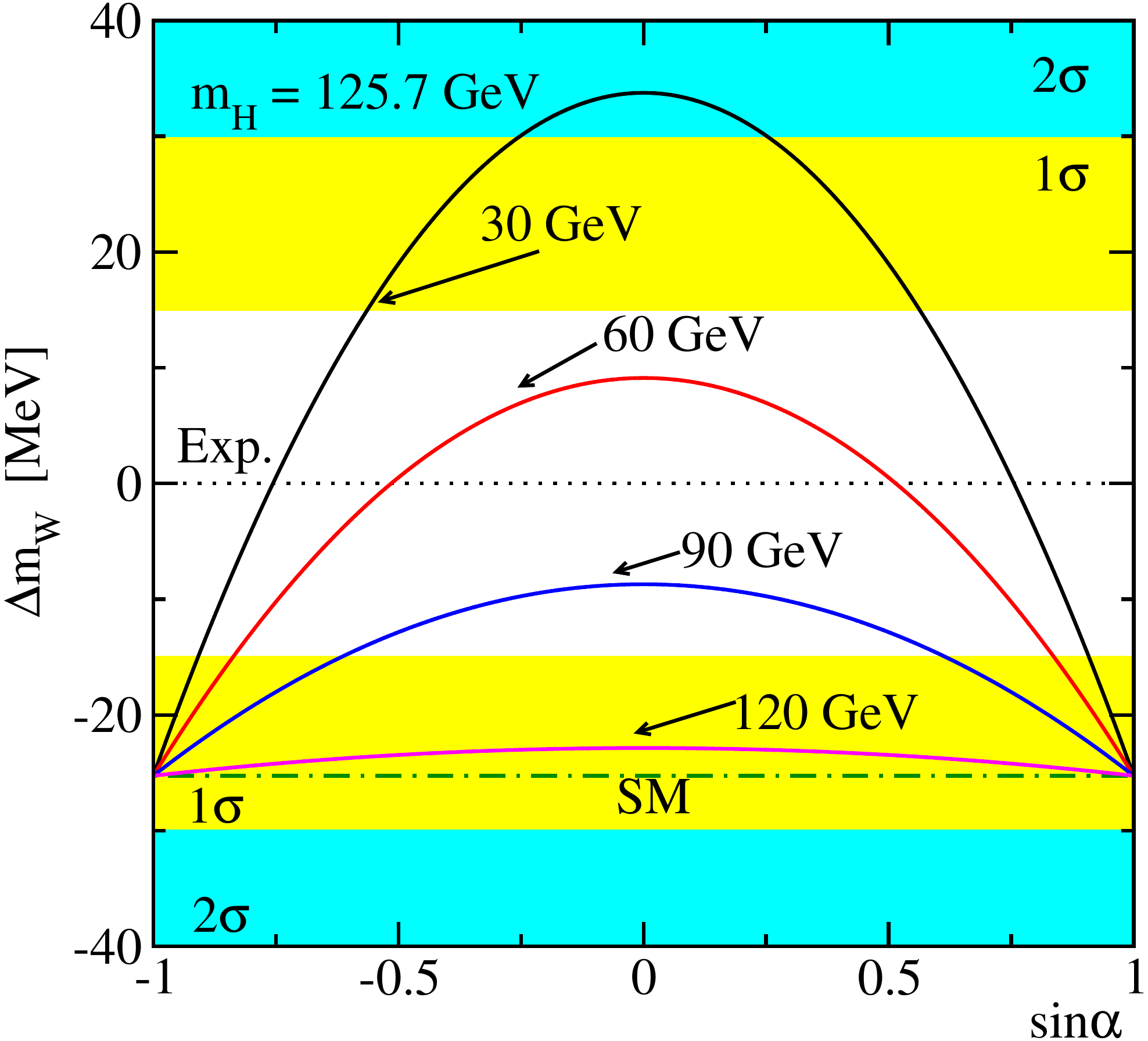} \\
\end{tabular}
\caption{{\underline{Upper panels}: full one--loop evaluation of {$\Delta r \equiv \Delta r_{\text{sing}}$} (left) and 
$\Delta m_W \equiv m_W^{\text{sing}}-m_W^\text{exp}$ (right) for different heavy Higgs masses [$m_{\Hzero}$]
with fixed [$m_{\hzero} = 125.7$ GeV], as a function
of the mixing angle [$\sin\alpha$]. \underline{Lower panels}: likewise,
for different light Higgs masses [$m_{\hzero}$] and fixed [$m_{\Hzero} = 125.7$ GeV].}
The corresponding SM predictions
(the experimental values) are displayed in dashed
(dotted) lines. The shaded bands
illustrate the $1\sigma$ and $2\sigma$ C.L. exclusion regions. 
Compatibility with the LHC signal strength measurements {requires} $|\sin\,\al|\,\lesssim\,0.42$ (upper panels) 
and $|\sin\al|\,\gtrsim\,0.91$ (lower panels) (c.f. section \ref{sec:comparison}). }
\label{fig:oversa}
\end{center}
\end{figure}

\begin{figure}[t!]
\begin{center}
\begin{tabular}{ccc}
\includegraphics[width=0.3\textwidth]{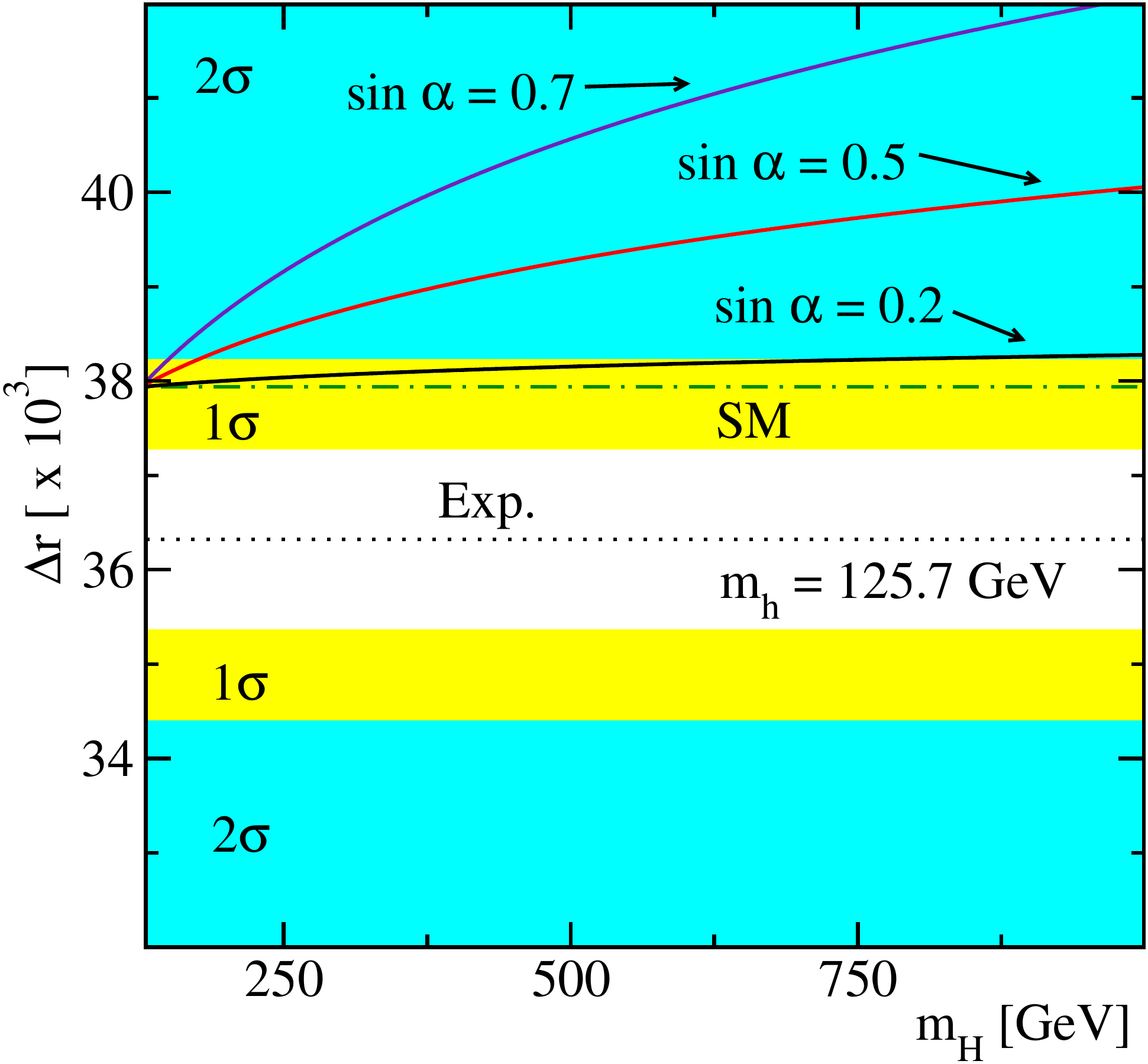} & & 
\includegraphics[width=0.31\textwidth]{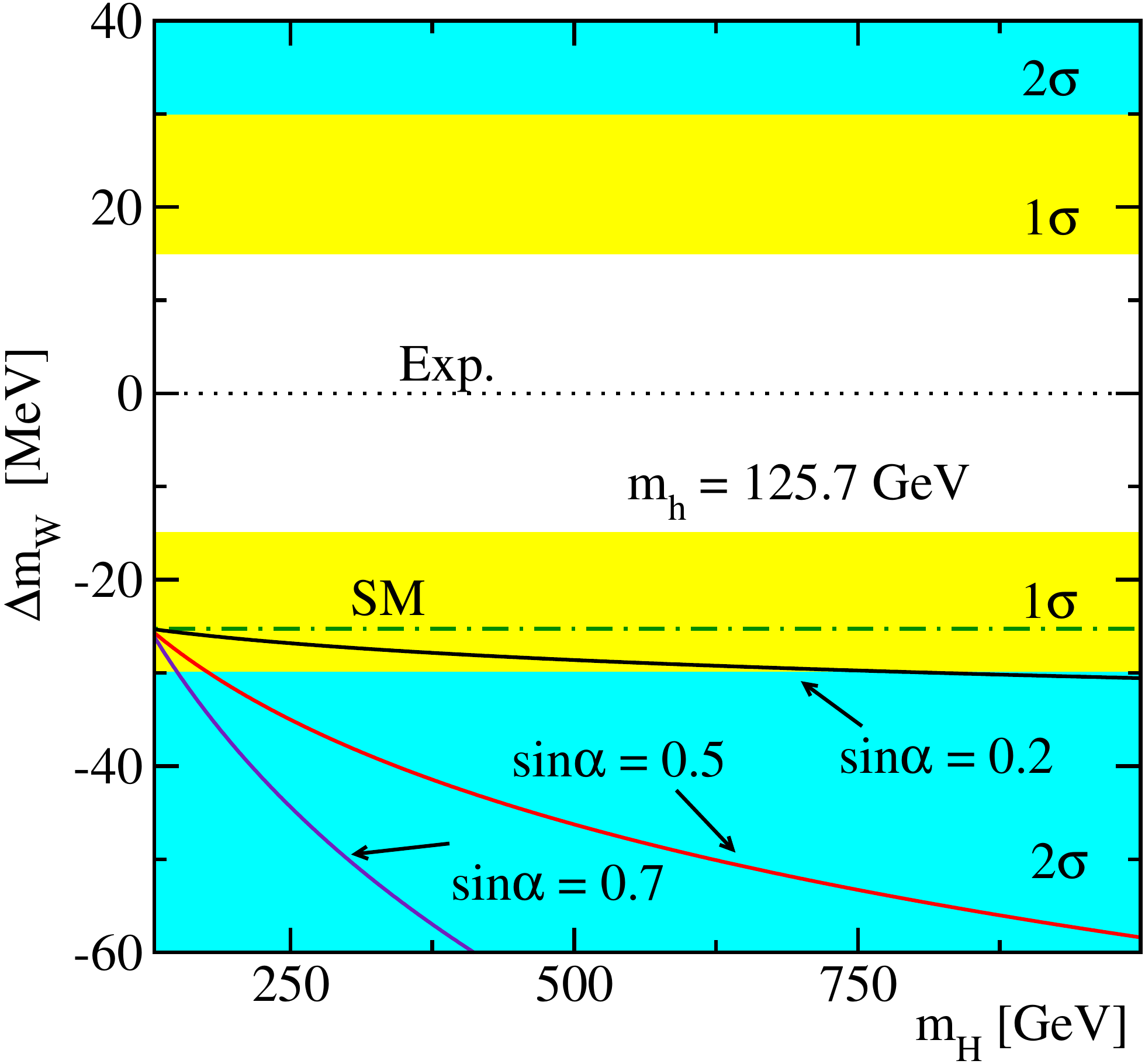} \\ & & \\
\includegraphics[width=0.3\textwidth]{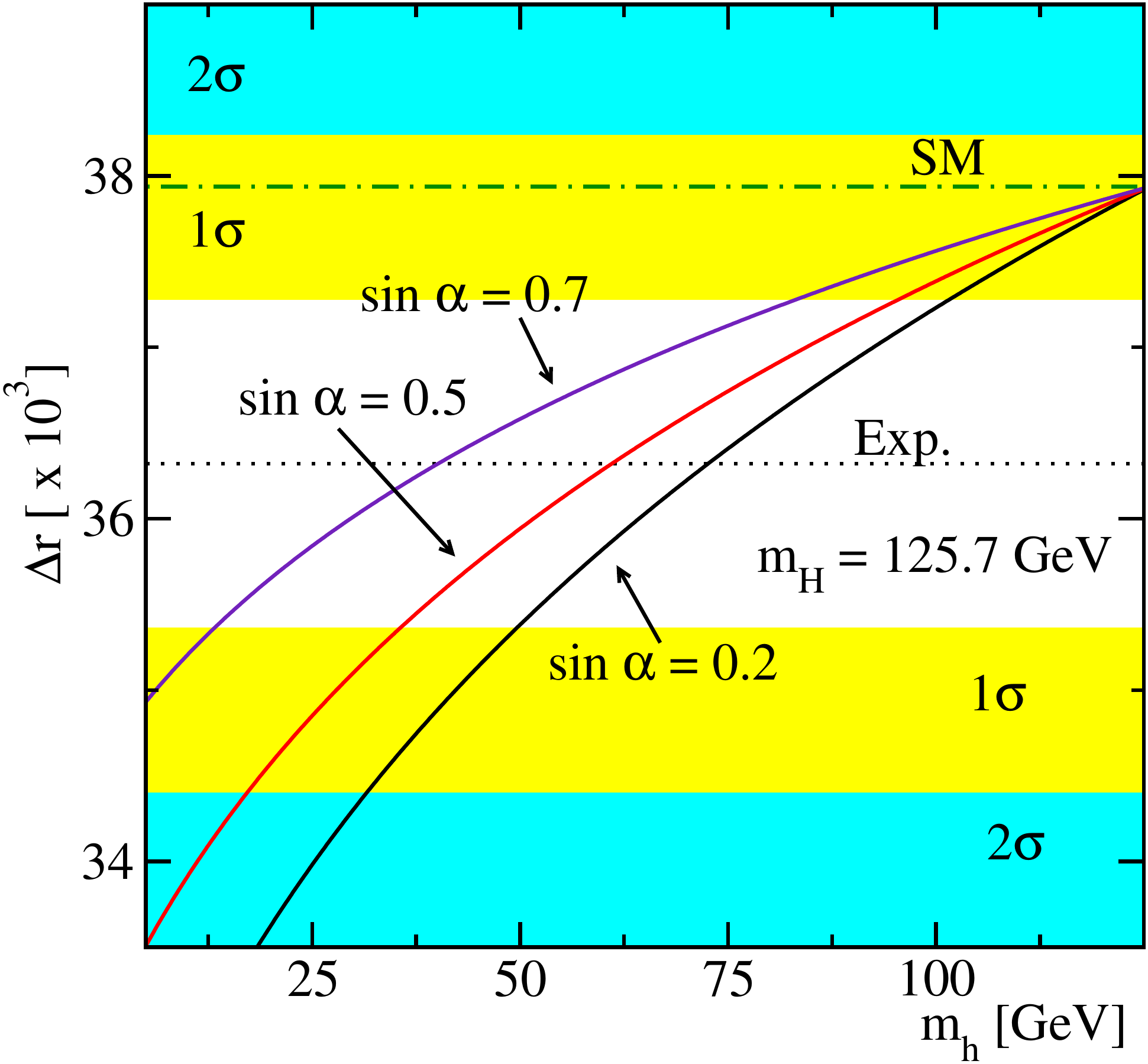} & & 
\includegraphics[width=0.31\textwidth]{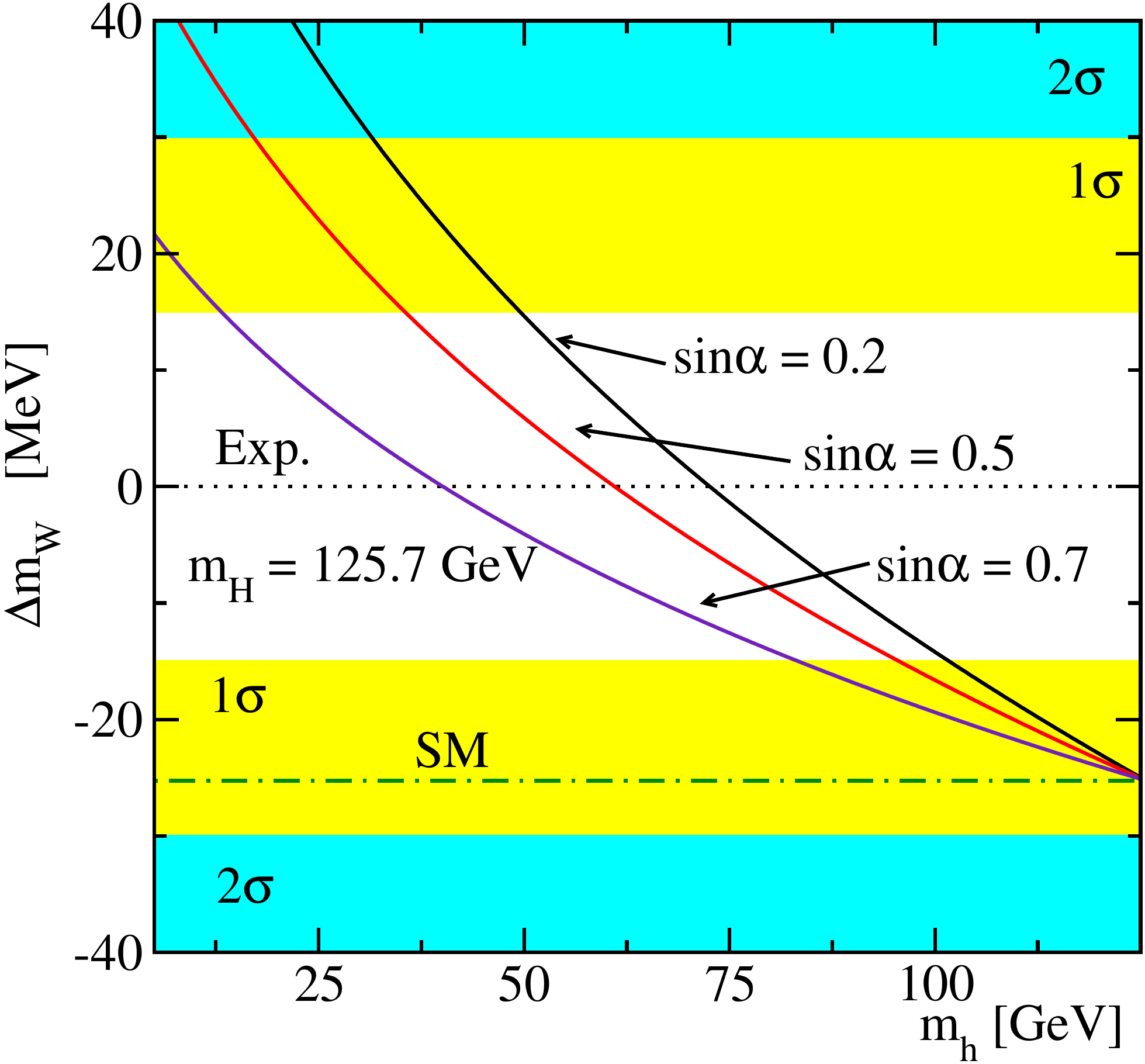} 
\end{tabular}
\caption{
Full one--loop evaluation of {$\Delta r \equiv \Delta r_{\text{sing}}$} (left) and 
$\Delta m_W \equiv m_W^{\text{sing}}-m_W^\text{exp}$ (right) for different mixing 
angle values, as a function of the heavy Higgs mass [$m_{\Hzero}$] 
{(upper panels) and the light Higgs mass [$m_{\hzero}$] (lower panels)}. 
The corresponding SM predictions
(the experimental values) are displayed in dashed
(dotted) lines. The shaded bands
illustrate the $1\sigma$ and $2\sigma$ C.L. exclusion regions.}
\label{fig:overmass}
\end{center}
\end{figure}

In the following we present an upshot of our numerical analysis.
Figures~\ref{fig:oversa} and \ref{fig:overmass} illustrate the behavior of 
{ $\Delta r \equiv \Delta r_{\text{sing}}$} and $\Delta m_W \equiv m^{\text{sing}}_W- m_W^\text{exp}$
under variations of the relevant singlet model
parameters. In
Figure~\ref{fig:oversa} 
we portray the evolution of both quantities
with the mixing angle, 
for illustrative Higgs companion masses. In the upper panels 
the SM--like Higgs particle is identified with the lightest singlet model mass--eigenstate [$\hzero$].
{We fix its mass to $m_{\hzero} = 125.7$ GeV} and
 sweep over a heavy Higgs mass range $m_{\Hzero} = 200 - 1000$ GeV.
The complementary case [$m_{\Hzero}= 125.7$ GeV $> m_{\hzero}$] is examined
in the lower panels, with a variable mass for the second {(light)} Higgs spanning $m_{\hzero} = 5 - 125$ GeV.
The results shown for $\Delta r$  
are referred to both the SM prediction [$\drsm$] and
the experimental value [$\drexp$]. The latter follows from
Eq.~\eqref{eq:deltar_def1} with the experimental inputs~\cite{Beringer:1900zz}
%

\begin{equation}
\begin{array}{ccc} m^{\rm exp}_{\PW} = 80.385 \pm 0.015\, \GeV \ & \; &
m_{\PZ} = 91.1876 \pm 0.0021 \,\GeV \\
 \alpha_{\text{em}}(0) = 1/137.035999074(44) &\; &  G_F = 1.1663787(6) \, 10^{-5} \,\GeV^{-2}  \label{eq:sminput}
\end{array},
\end{equation}

\noindent wherefrom we get

\begin{alignat}{5}
 \Delta r_{\rm exp} &= \frac{\sqrt{2}\,G_F}{\pi\alpha_{\text{em}}}\,m_{\PW}^2\,\left(1-\frac{m_{\PW}^2}{m_Z^2}\right) - 1 = \lb \right. 36.320 \left.\,\pm\,0.976\rb \times 10^{-3}
\label{eq:deltarexp}.
\end{alignat}

\noindent The 1$\sigma$ and 2$\sigma$ C.L. regions in $\Delta r_{\text{exp}}$ are derived from the $m_W^{\text{exp}}$ uncertainty
bands using standard error propagation.

\smallskip{}{
Figure~\ref{fig:overmass} provides a complementary view} of the $\Delta r$ and $\Delta m_W$ dependence on
the additional Higgs boson mass
assuming mild ($\sin \alpha = 0.2$), moderate ($\sin \alpha = 0.5$)
and strong ($\sin \alpha = 0.7$) mixing.

\medskip{}
These plots nicely illustrate the parameter dependences anticipated earlier {e.g. in Eqs.~\eqref{eq:selfzz}-\eqref{eq:selfww}}. 
On the one hand, the quadratic $\sin^2\alpha$ ($\cos^2\alpha$) 
dependence
reflects the global rescaling of the light (heavy) SM--like Higgs coupling
to the weak gauge bosons. Accordingly, the values of $\Delta r$ and {$m^{\text{sing}}_W$}
converge to the SM predictions in the limit $\sin\alpha = 0$ ($\sin\al = \pm 1$)
in which the new physics effects decouple. The
growing departure from the SM as we raise (lower) the mass of the heavy (ligher) Higgs
companion follows the {logarithmic} behavior singled out in Eq.~\eqref{eq:deltarho-singlet2},
{and can be traced back to the increasing breaking of the {(approximate)} custodial invariance.}

\medskip{}
In the case where $m_{\hzero}\,=\,125.7\,\GeV$ and $m_{\Hzero}>130\,\GeV$ 
(cf. upper panels of
Figs. \ref{fig:oversa} and \ref{fig:overmass}), we pin down positive (negative) deviations
of $\drsing$ ($m^{\text{sing}}_W$) with
respect to the corresponding SM predictions. These increase
systematically for larger mixing angles and heavier Higgs companions. The stark dependence
on $\sin\alpha$ and $m_{\Hzero}$, combined with 
the fact that $m^{\text{sing}}_W - m_W^{\text{SM}} < 0$ and that $m_W^{\text{SM}}$ already
lies 20 GeV below the experimental measurement, explains why the results obtained in this
case can easily lie outside of the $2 \sigma$ C.L. exclusion region.
In the complementary scenario (cf. lower pannels),
in which we set $m_{\Hzero}\,=\,125.7\,\GeV$ and vary 
the light Higgs mass $m_{\hzero}\,\leq\,125\,\GeV$, we find analogous trends 
-- but with interchanged dependences. 
Here the additional one--loop effects from
the light Higgs companion help to release the $m_W^{\text{sing}}-m_W^{\text{exp}}$
tension. 
On the other hand, the onset of $2\sigma$--level constraints appears for $m_{\hzero} \lesssim 30$ GeV.
These results spotlight
a significant mass range in which
the singlet model contributions
could in principle achieve $m_W^{\text{sing}} \simeq m_W^{\text{exp}}$\footnote{
Let us recall that both instances $m^{\text{{th}}}_W - m^{\text{exp}}_W \lessgtr 0$ 
are possible in the 2HDM for a large variety of Higgs mass spectra. However, unlike
the singlet model case, these situations are not attached
to a specific mass hierarchy ~\cite{LopezVal:2012zb}.}.
{The viability of these scenarios is nevertheless hindered in practice,
due to the direct collider mass bounds and 
the LHC signal strength measurements. The impact of these additional constraints, which
at this point we have not yet included, will be {addressed} in section~\ref{sec:comparison}} \footnote{
A fully comprehensive analysis of the model {combining} all {currently available} constraints
deserves a dedicated study and will be presented elsewhere\cite{tt}.} 

\medskip{}

\begin{table}[tb!]
\begin{center}
 \begin{tabular}{|c|ccc|ccc|ccc|} \hline
  & \multicolumn{3}{|l|}{$\Delta r_{\text{sing}}\,[\times 10^3]$} & \multicolumn{3}{|l|}{$m^{\text{sing}}_{\PW}-m_{\PW}^{\text{exp}}$ [MeV]} 
  & \multicolumn{3}{|l|}{$\Delta(\delta\rho_{\text{sing}}) \; [\times 10^4]$} 
   \\ \hline \hline   \multicolumn{10}{|c|}{\textbf{$\hzero$ SM--like} [$m_{\hzero} = 125.7$ GeV]} \\ \hline

$m_{\Hzero}$ [GeV]  & 300 & 500 & 1000  & 300 & 500 & 1000 & 300 & 500 & 1000 \\ \hline  \hline
$\sin\alpha = 0.2$ & 38.067 & 38.153 & 38.277 & -27 & -29 & 
-31 & 
-0.241 & -0.428 & -0.711 \\ 
$\sin\alpha = 0.5$ & 38.744 & 39.281 & 40.056 & -38 & -46 & -58
& -1.508 & -2.674 & -4.450 \\ 
$\sin\alpha = 0.7$ & 39.515 & 40.565 & 42.077 & -50 & -66 & -90
& -2.956 & -5.244 & -8.730 \\ \hline
\hline   \multicolumn{10}{|c|}{\textbf{$\Hzero$ SM--like} [$m_{\Hzero} = 125.7$ GeV]} \\ \hline
$m_{\hzero}$ [GeV]  & 30 & 60 & 90  & 30 & 60 & 90 & 30 & 60 & 90 \\ \hline  \hline
$\sin\alpha = 0.2$ & 34.305 & 35.824 & 36.921 & 31 & 8 & -9 & 3.798 & 2.707 & 1.466 \\
$\sin\alpha = 0.5$ & 35.103 & 36.288 & 37.144 & 19 & 1 & -13 & 2.968 & 2.115 & 1.146 \\
$\sin\alpha = 0.7$ & 36.012 & 36.816 & 37.398 & 5 & -8 & -17 & 2.019 & 1.439 & 0.779 \\ \hline
 \end{tabular}
 \end{center}
 \caption{Parameter space survey of the electroweak parameters 
 $\Delta r_{\text{sing}}$, $\Delta m_W \equiv m_W^{\text{sing}} - m_W^{\text{{exp}}}$ and $\delta\rho \equiv \Delta(\delta\rho_{\text{sing}})$ 
 in the singlet model for representative Higgs masses and mixing angle choices. }
 \label{tab:results}
\end{table}

Our discussion is complemented by specific numerical predictions for $\Delta r$ and {$m^{\text{sing}}_W$}$-m_W^\text{exp}$, 
{which we list}  in Table~\ref{tab:results} for representative
parameter choices. 
For small mixing $|\sin\alpha| \lesssim 0.2$ and heavy Higgs masses of few hundred GeV,
$\Delta r_{\text{sing}}$ {departs from} $\drsm$
at the $\mathcal{O}(0.1)\%$ level.
These {deviations} may increase up to $\mathcal{O}(10)\%$ for mixing angles above $|\sin\alpha| \gtrsim 0.5$ 
and $\mathcal{O}(1)$ TeV scalar companions. {Not surprisingly,} these are the parameter space configurations
that maximize the non--standard singlet model imprints, viz. {the rescaled Higgs boson interactions} and the non--decoupling mass dependence
of the {Higgs--mediated loops}. {As we have seen in Figures~\ref{fig:oversa}-\ref{fig:overmass}, and according 
to Eq.~\eqref{eq:mwshift}},
these shifts pull the resulting prediction [$m^{\text{sing}}_W$]  down to
$\sim 1 - 70$ MeV below
the SM result. 
Staying within $1\sigma$ C.L.
we find $|m^{\text{{sing}}}_W - m_W^{\text{SM}}| \simeq 10 \,\MeV$ ($m^{\text{sing}}_W < m_W^{\text{SM}}$) 
for relatively tempered mixing ($|\sin\alpha| \lesssim 0.2$) and
heavy Higgs masses up to 1 TeV. {Larger mixings of typically $|\sin\alpha| \gtrsim 0.4$
push $m_{W}^{\text{sing}}$ into the $2\sigma$--level exclusion region.}
These results once more illustrate that,
for a second {\sl heavy} Higgs resonance, the singlet model effects
tend to sharpen {the $m_W^{\text{th}}-m_W^{\text{exp}}$ tension} even further,
and more so as we increasingly depart from the SM--like limit.
Alternatively, for $m_{\hzero} < m_{\Hzero} = 125.7$ GeV
we find that relative deviations of $\sim 5\%$ in $\drsing$
( with $\drsing < \drsm$) are attainable for $50-100$ GeV light Higgs companion masses
and mixing angles above $|\sin\alpha| \sim 0.5$. 
\begin{figure}[t!]
\begin{center}
\begin{tabular}{ccc}
\includegraphics[width=0.29\textwidth]{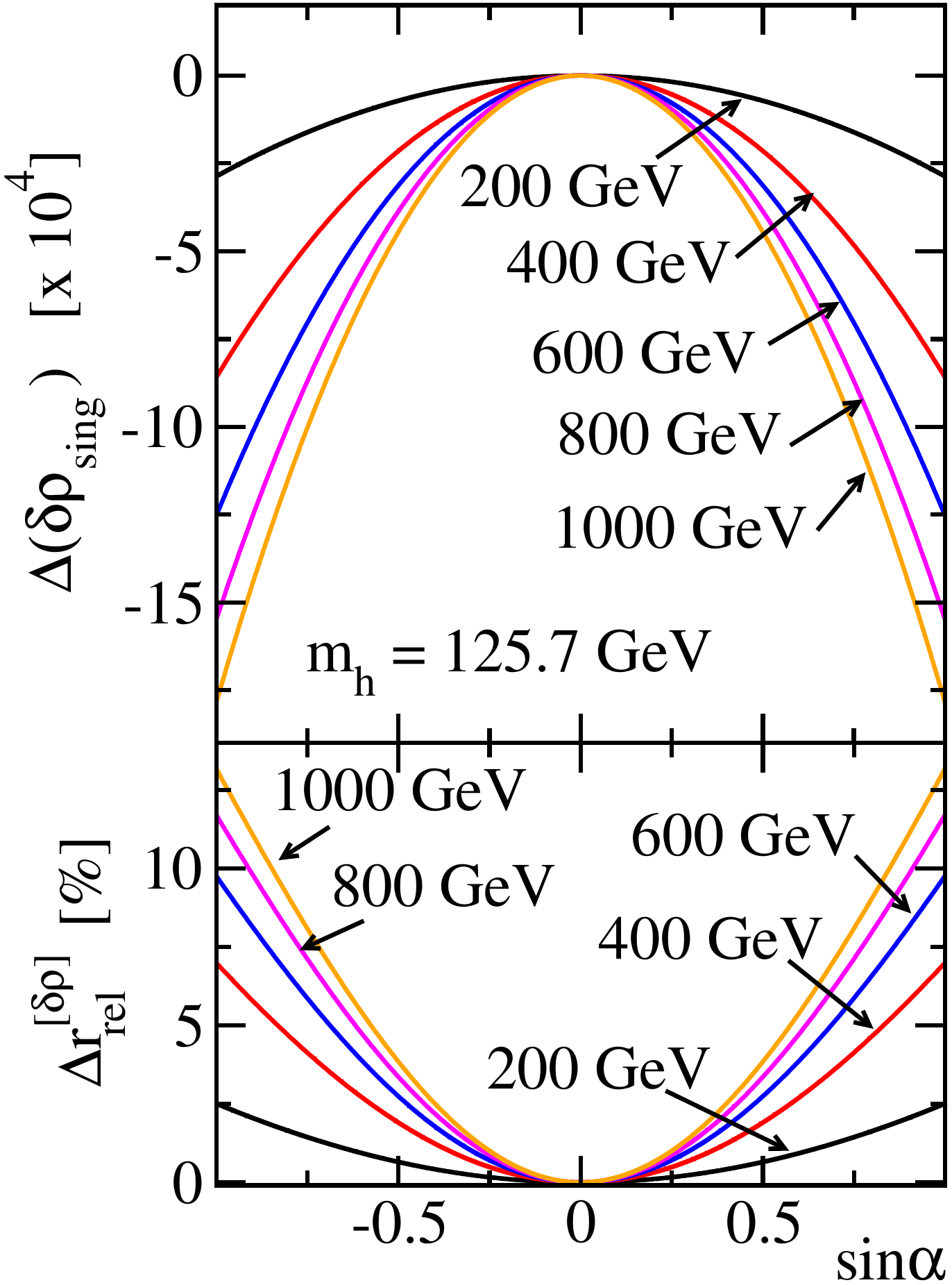} & & 
\includegraphics[width=0.28\textwidth]{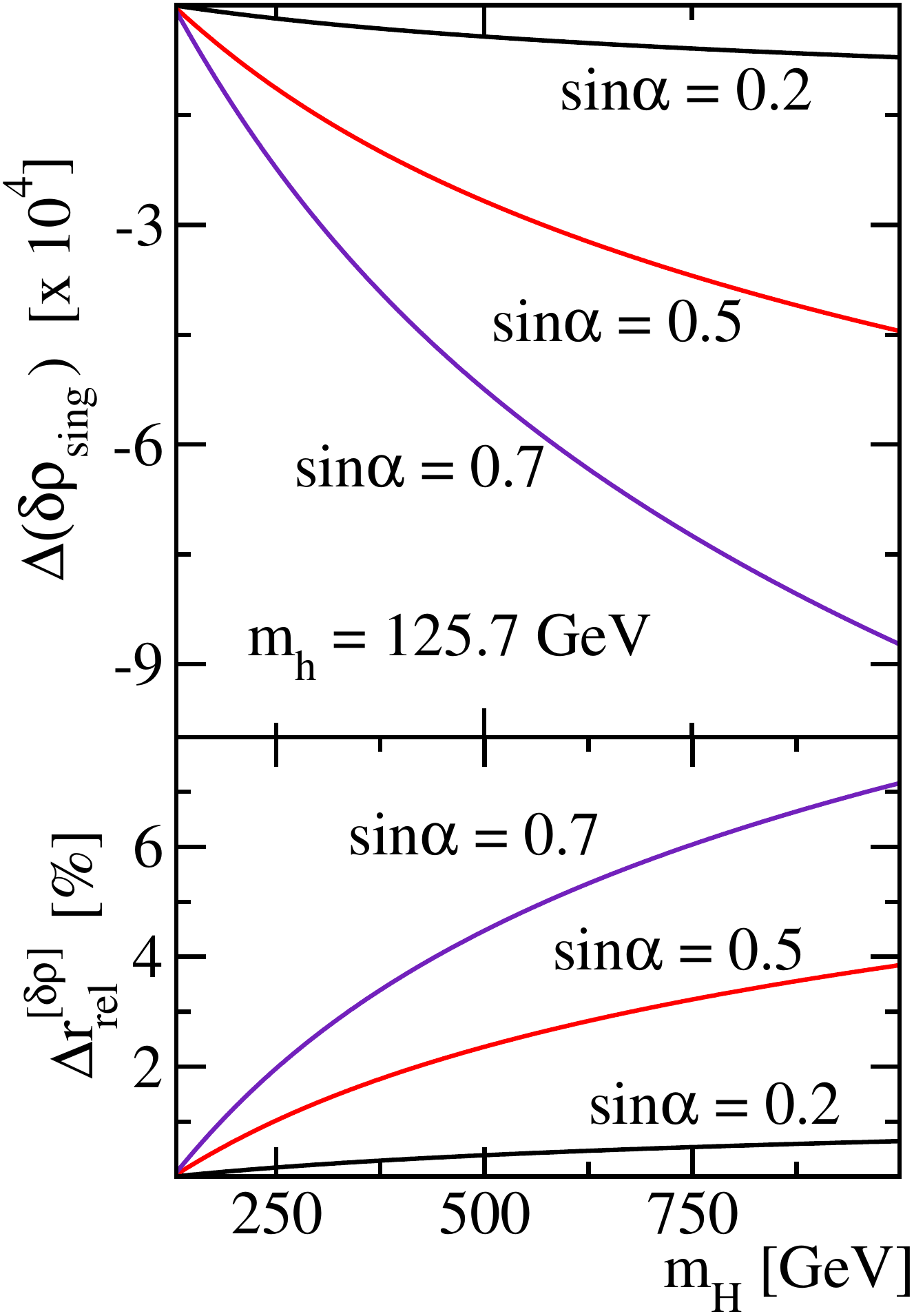} \\ & & \\
\includegraphics[width=0.29\textwidth]{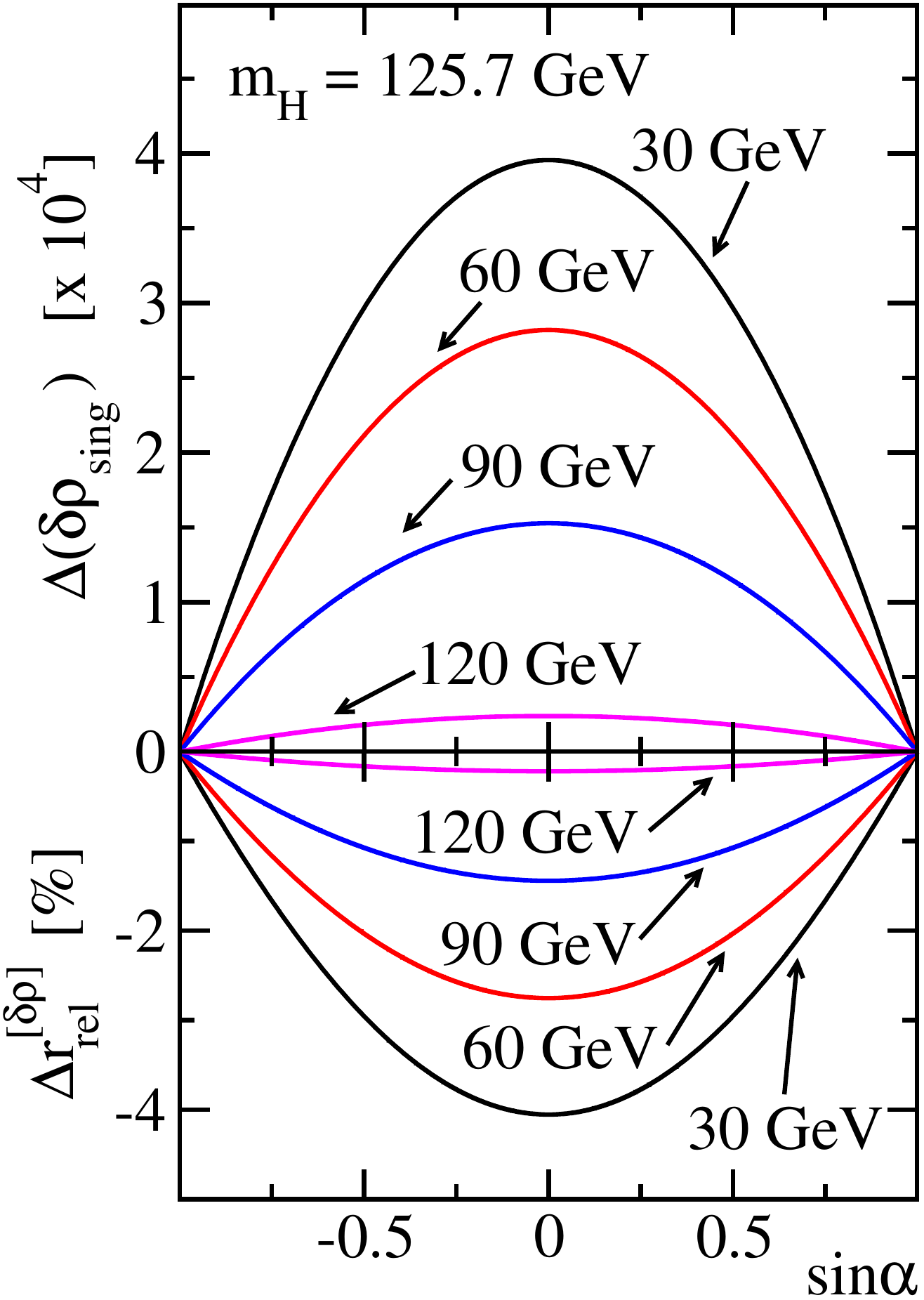} & & 
\includegraphics[width=0.28\textwidth]{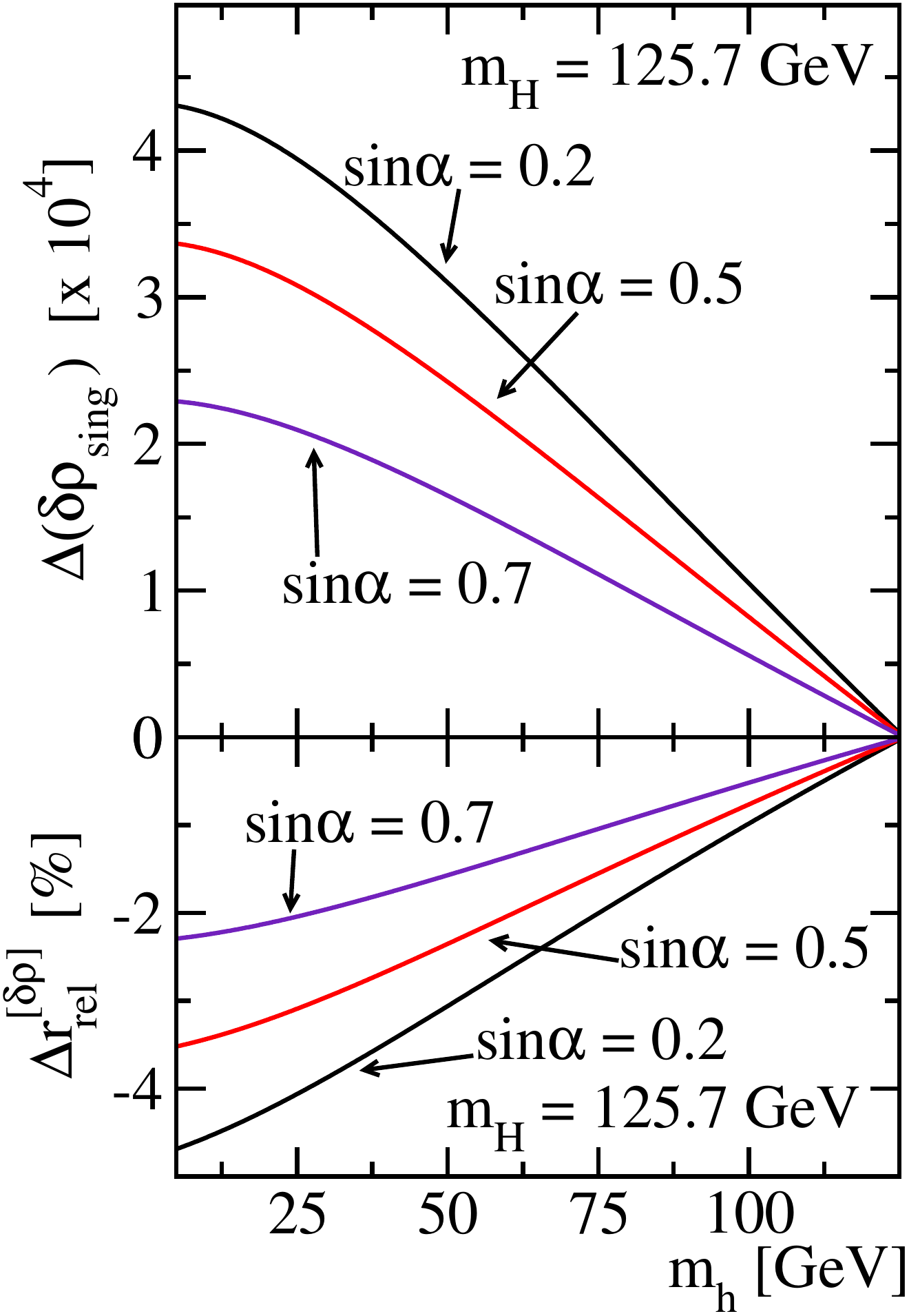} 
\end{tabular}
\caption{Singlet model contribution to the $\delta\rho$ parameter
at one loop [$\ddeltarhosing$]
for representative
mixing angles {and Higgs companion masses}. 
In the bottom subpannels we quantify the relative size of these
contributions
to the overall $\Delta r$ prediction, following Eq.~\eqref{eq:ratio}. }
\label{fig:deltarho}
\end{center}
\end{figure}

\bigskip{}

Alongside with the calculation of $\Delta r$ and {$m^{\text{sing}}_W$} 
we compute the new physics one--loop contributions to the $\delta\rho$ 
parameter (cf. Eq.~(\ref{eq:deltarho-def1})).
The behavior of $\ddeltarhosing$ as a function of the relevant singlet model parameters
 $[\sin\alpha]$ and $[m_{\hzero,\Hzero}]$ is illustrated in Figure~\ref{fig:deltarho}. 
{Again, we separately
examine the two complementary situations in which either the ligher (top--row panels)
or the heavier (bottom--row panels) singlet model mass--eigenstate 
describes the SM--like Higgs boson. 
As expected, $|\ddeltarhosing|$ enlarges as we progressively separate from the SM limit. The 
strong dependence in the {additional} Higgs mass displays the increasing deviation from
the custodial symmetry limit, {which is enhanced by} the mass splitting 
between the Higgs mass--eigenstates.  
{Conversely, we recover $\ddeltarhosing \to 0$ in the $m_{\hzero} \to m_{\Hzero}$ limit.} 
{The relative size of the static one--loop effects
encapsulated in $\ddeltarhosing$}
is quantified in the lower subpannels 
of Fig.~\ref{fig:deltarho} through the ratio
\begin{equation}
 \Delta r_{\text{rel}}^{[\delta\rho]} \equiv \Delta r_{\text{sing}}^{[\delta\rho]}/\drsing = -\cwd/\swd\,\ddeltarhosing / \drsing,
 \label{eq:ratio}
 \end{equation}
\noindent  which we construct from the different pieces singled out in Eq.~\eqref{eq:deltar_def4}, 
retaining the singlet model contributions only.

\medskip{} 
Interestingly, the analysis of $\delta\rho$
provides a handle for estimating the size 
of higher--order corrections. 
The leading singlet model two--loop effects  [$\Delta(\delta\rho_{\text{sing}}^{[2]})$]
arise from the exchange of virtual top quarks and Higgs bosons. 
This type of mixed $\mathcal{O}(G_F^2 m_t^4)$ Yukawa corrections
was first computed within the SM
in the small Higgs boson mass limit in Ref.~\cite{vanderBij:1986hy} and
later on extended to arbitrary masses~\cite{Barbieri:1992dq,Fleischer:1993ub}. 
The analytical expressions therewith can be readily exported to our case. 
Taking into account the {rescaled}
top--quark interactions with the 
light (heavy) Higgs mass--eigenstate by an overall factor $\sim \cos^2\alpha$ ($\sim \sin^2\alpha$); and removing as usual the overlap with the SM contribution
(which we identify here with $\hzero$ in the $\sin\alpha = 0$ limit)
we find 

\begin{eqnarray}
\lefteqn{ \Delta(\delta\rho_{\text{sing}}^{[2]}) =}\\
&& \cfrac{{3}\,G_F^2\,m_t^4\,\sin^2\alpha}{128\,\pi^4}\,
 \left( f(m^2_t/m^2_{\Hzero})  - f(m^2_t/m^2_{\hzero})\right) 
   \sim \cfrac{{3\,}G_F^2\,m_t^4\,\sin^2\alpha}{128\,\pi^4}\,
 \left[ 27\,\log\left(\cfrac{m_t}{m_{\Hzero}}\right) + \cfrac{4\pi m_{\hzero}}{m_t}\right],\nonumber
 \label{eq:ndeltarhotop}
\end{eqnarray}

\noindent where in the latter step we have introduced
the asymptotic expansions of $f(r)$ ~\cite{Barbieri:1992dq,Fleischer:1993ub}.
The above estimate $\Delta(\delta\rho_{\text{sing}}^{[2]})$
stagnates around $\mathcal{O}(10^{-4})$
for fiducial parameter choices with $|\sin\al|\lesssim 0.5$. 
 When 
promoted to the W--boson mass prediction through   
Eqs.~\eqref{eq:mwshift}
and \eqref{eq:deltar_def4} we find

\begin{alignat}{5}
\Big{[} \Delta(m_W^{[2]})\Big{]}_{\text{sing}} \sim -\cfrac{1}{2}\,m_W\,\cfrac{\swd}{\cwd-\swd}\,\delta(\Delta r_{\text{sing}}^{\delta[\rho]})
 \lesssim \mathcal{O}(1) \text{MeV} \label{eq:mwhigher},
\end{alignat}

\noindent which we can interpreted as an estimate on the theoretical
uncertainty on $m_{W}^{\text{sing}}$ due to the quantum effects beyond the one--loop order.

\begin{table}[t!]
\begin{\eqn*}
\begin{array}{c|c}
m_{H}\,[\GeV]&|\sin\al|_{\text{max}}\\ \hline\hline
1000&0.19\\
900&0.20\\
800&0.20\\
700&0.21\\
600&0.22\\
500&0.24\\
400&0.26 \\
300&0.31\\
200&0.43\\ 
150&0.70\\
130&1.00
\end{array}
\end{\eqn*}
\caption{\label{tab:vals} 
{Upper limits on the} mixing angle compatible with $m^{\text{exp}}_W$ at the $2\sigma$--level, 
for $m_{\hzero} = 125.7$ GeV and representative heavy Higgs masses. Consistency with the
 LHC signal strength measurement implies 
 $|\sin\al|\,\leq\,0.42$, cf. Fig. \ref{fig:expbounds}.
}
\end{table}

\subsection{Comparison to complementary model constraints}

\label{sec:comparison}

In this section, we first {confront} 
the model constraints imposed by the [$m_W^{\text{sing}} - m_W^{\text{exp}}$] comparison to those following from
global fits to electroweak precision data. {The difference [$m_W^{\text{sing}} - m_W^{\text{exp}}$]  
corresponds to a
(pseudo)observable which can directly be linked to a single experimental measurement}.
The electroweak precision tests are customary} expressed
in terms of the oblique parameters $[S,T,U]$, {c.f. e.g. Refs.~}
\cite{BahatTreidel:2006kx,Bowen:2007ia,Bhattacharyya:2007pb,Dawson:2009yx,Englert:2011yb,Gupta:2012mi,Batell:2012mj} {for analyses of the singlet extension with a $\mathcal{Z}_2$ symmetry}, 
and \cite{Profumo:2007wc,Cline:2009sn} for a slightly different model setup. In the {standard} conventions \cite{Beringer:1900zz}, and retaining
the one--loop singlet model {contributions} only, {these parameters are given by } 

\begin{eqnarray}
\cfrac{\alpha_{\text{{em}}}}{4\swd\cwd}\,S &=& 
\cfrac{\overline{\Sigma}_{\PZ}(m_Z^2) - \overline{\Sigma}_{\PZ}(0)}{m_{\PZ}^2}; 
\qquad
 \alpha_{\text{{em}}}\,T = \cfrac{\overline{\Sigma}_{\PW}(0)}{m_{\PW}^2} -
\cfrac{\overline{\Sigma}_{\PZ}(0)}{m_{\PZ}^2}; \nonumber \\
 \cfrac{\alpha_{\text{{em}}}}{4\swd}\,U &=& \cfrac{\overline{\Sigma}_{\PW}(m^2_W) -
\overline{\Sigma}_{\PW}(0)}{m_W^2} - 
\cwd\,\cfrac{\overline{\Sigma}_{\PZ}(m^2_Z) -
\overline{\Sigma}_{\PZ}(0)}{m_Z^2}.
\label{eq:oblique}
\end{eqnarray}

\noindent Notice that {genuine singlet model} contributions to
the photon and the mixed photon--$\PZ$ vacuum polarization {are absent at one loop}. 
The overlined notation $\overline{\Sigma}$ is once
more tracking down
the consistent subtraction of the overlap with
the SM Higgs--mediated contributions, as specified by Eq.~\eqref{eq:subtractoverlap}.
The $T$ parameter can obviously be related to the $\delta\rho$ parameter
in Eq.~\eqref{eq:deltarho-def1},
{yielding} $\alpha_{{\text{em}}}\,T = -\delta\rho$. 
Likewise, we may rewrite 
$\Delta r_{\text{sing}}$  as

\begin{equation}
 \Delta r_{\text{sing}} = \cfrac{\alpha_{\text{{em}}}}{\swd}\,\left(-\cfrac{1}{2}\,S + \cwd\,T + \cfrac{\cwd-\swd}{4\swd}\,U \right)
 \label{eq:deltar-oblique}.
\end{equation}

In Fig.~\ref{fig:oblique} we {portray the functional dependence} $[S,T,U]$ {with respect to}
the {relevant singlet model parameters}.
The best--fit point has been taken from Ref.~\cite{Baak:2012kk},
 including the 
LHC Higgs mass measurement of
  $126.7\,\pm\,0.4\,\GeV$ as an input parameter,
and yields

\begin{alignat}{5}
 S = 0.03 \pm 0.10; \qquad T = 0.05 \pm 0.12; \qquad U = 0.03 \pm 0.10 
 \label{eq:bestfit-oblique}.
\end{alignat}

{Correlations among these parameters are revelant and must be taken into account when 
{electroweak precision} global fit estimates are used to constrain the parameter space of the model. {To that aim}
we here use the best linear unbiased estimator (see e.g. \cite{Lyons:1988rp}) based on the
 Gauss--Markov theorem
which yields}

\begin{alignat}{5}
 \chi^2 &= \,(\mathcal{O}_l - \mathcal{\hat{O}}_l)\,(V_{lk})^{-1}\,(\mathcal{O}_k - \mathcal{\hat{O}}_k)\qquad \text{with} \qquad \mathcal{O}_l = \{S,T,U\}
 \label{eq:chicorrel},
\end{alignat}

\noindent where $\mathcal{\hat{O}}_l$ stand for the
global best--fit values {of} the oblique parameters in Eq.~\eqref{eq:bestfit-oblique}.
The covariance matrix $V_{lk}$ is extracted from
Ref.~\cite{Baak:2012kk}, with correlation coefficients between
{the parameter pairs  [$(S,T),\,(S,U),\,(T,U)$] {given by [+0.89, -0.54, -0.83]} respectively.
} 

{We carry out our analysis by 
fixing the heavy (resp. light) additional scalar mass and 
allowing for correlated variations of up to $2\sigma$ in each of these parameters. 
For a two--parameter estimate, this {translates into} $|\Delta\,\chi| \leq 5.99$. 
That way 
we derive upper (resp. lower) mixing angle limits,
which correspond to the parameter space regions compatible with these global
electroweak precision tests. 
{Albeit
rendering non-negligible contraints, we find the resulting limits (c.f. e.g. the magenta line of Fig.~\ref{fig:expbounds}) to be superseded by other constraints throughout the entire
 parameter space,
  as we discuss below.}
\medskip{}

\begin{figure}[t!]
\begin{center}
 \begin{tabular}{ccc}
 \includegraphics[width=0.28\textwidth]{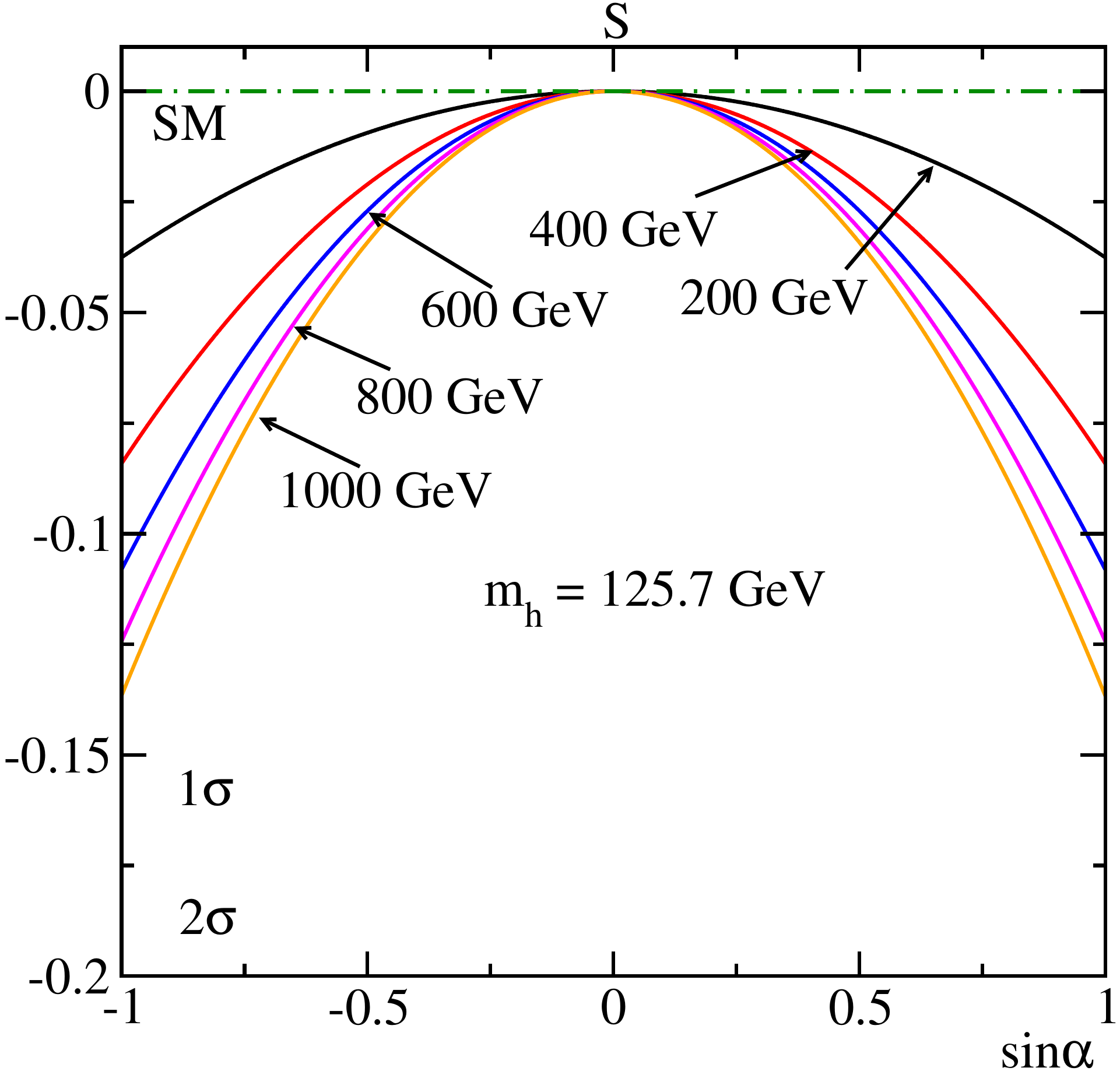} & \includegraphics[width=0.28\textwidth]{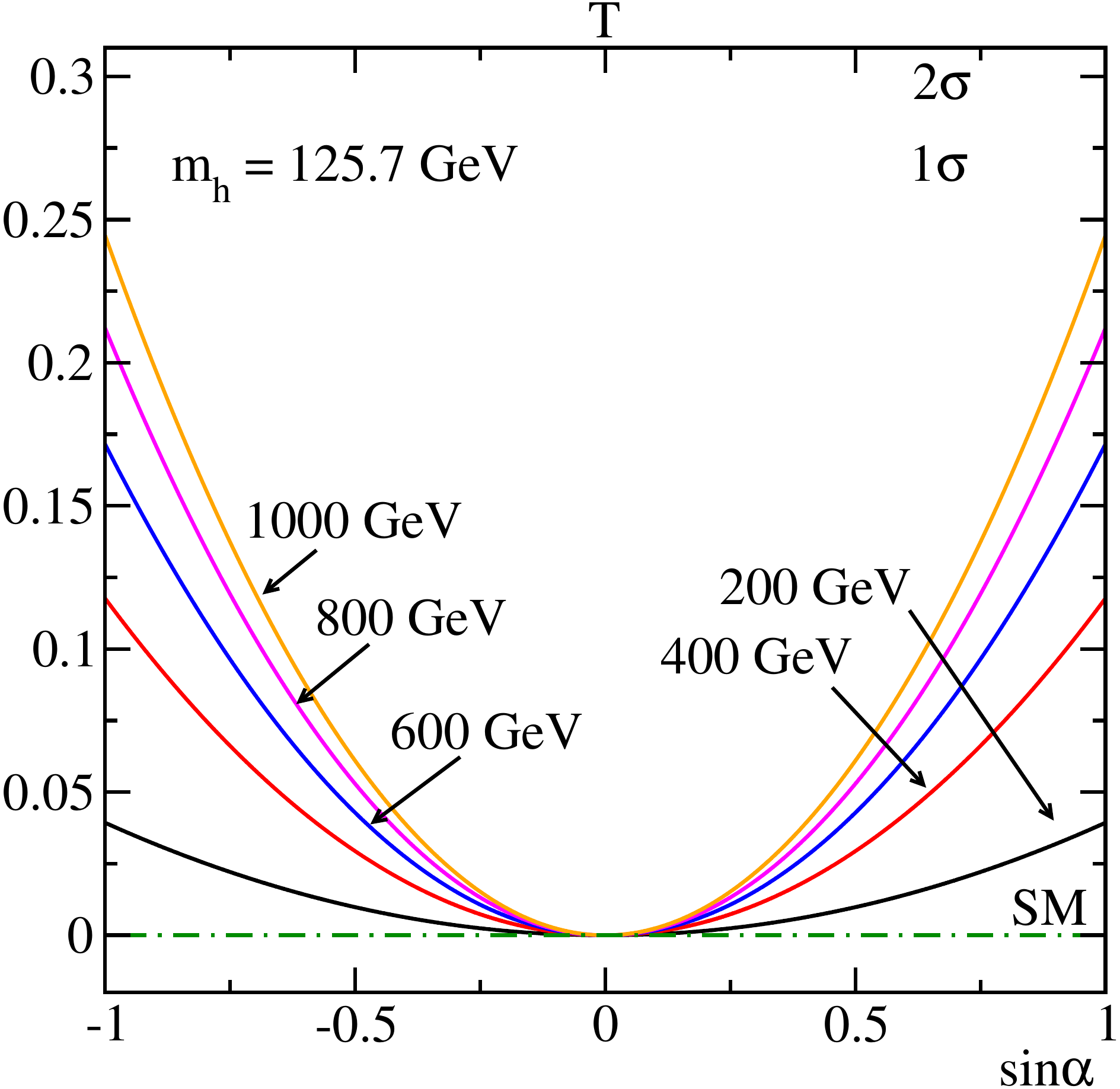} &  \includegraphics[width=0.28\textwidth]{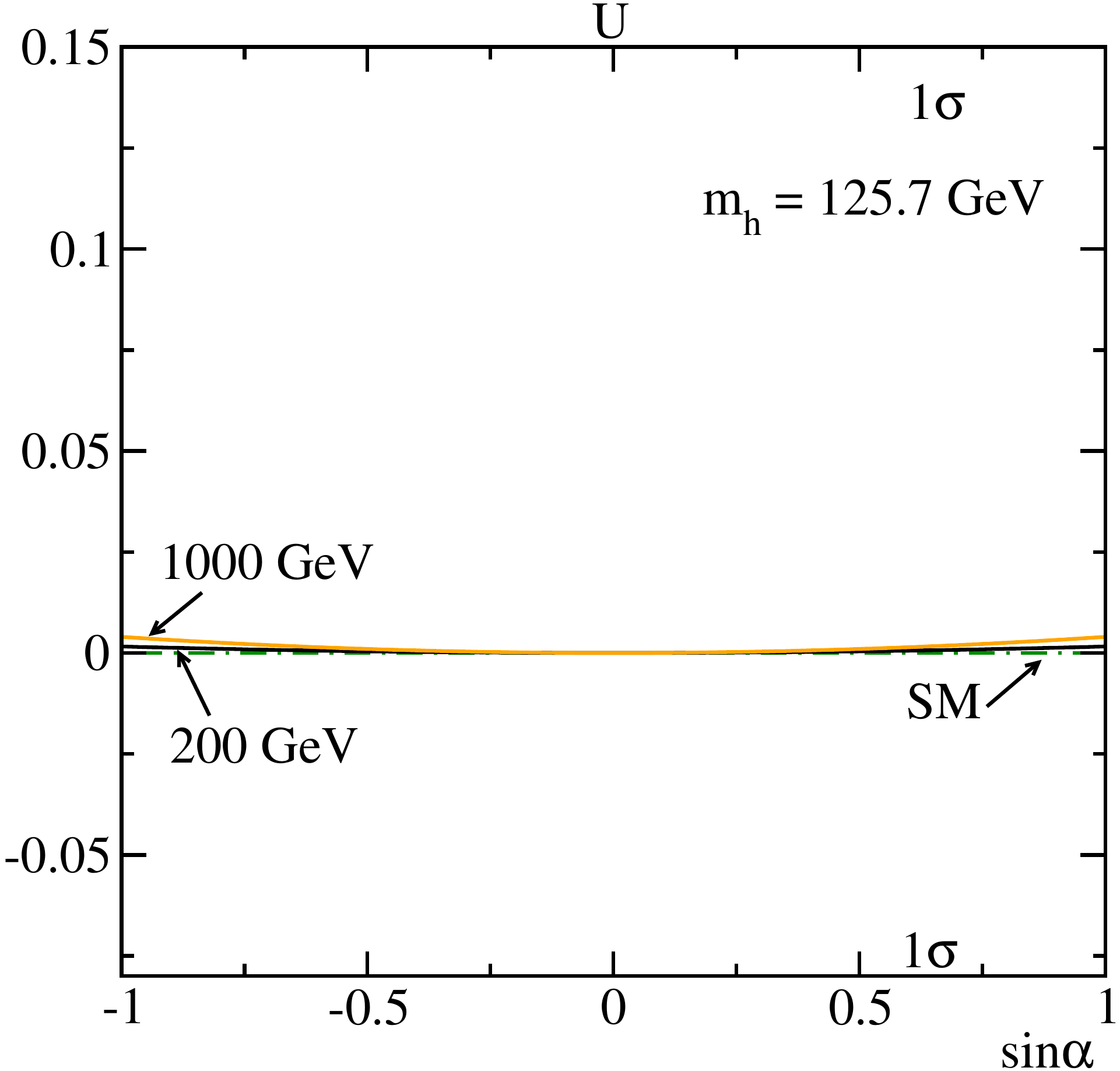} \\
 & & \\
 \includegraphics[width=0.28\textwidth]{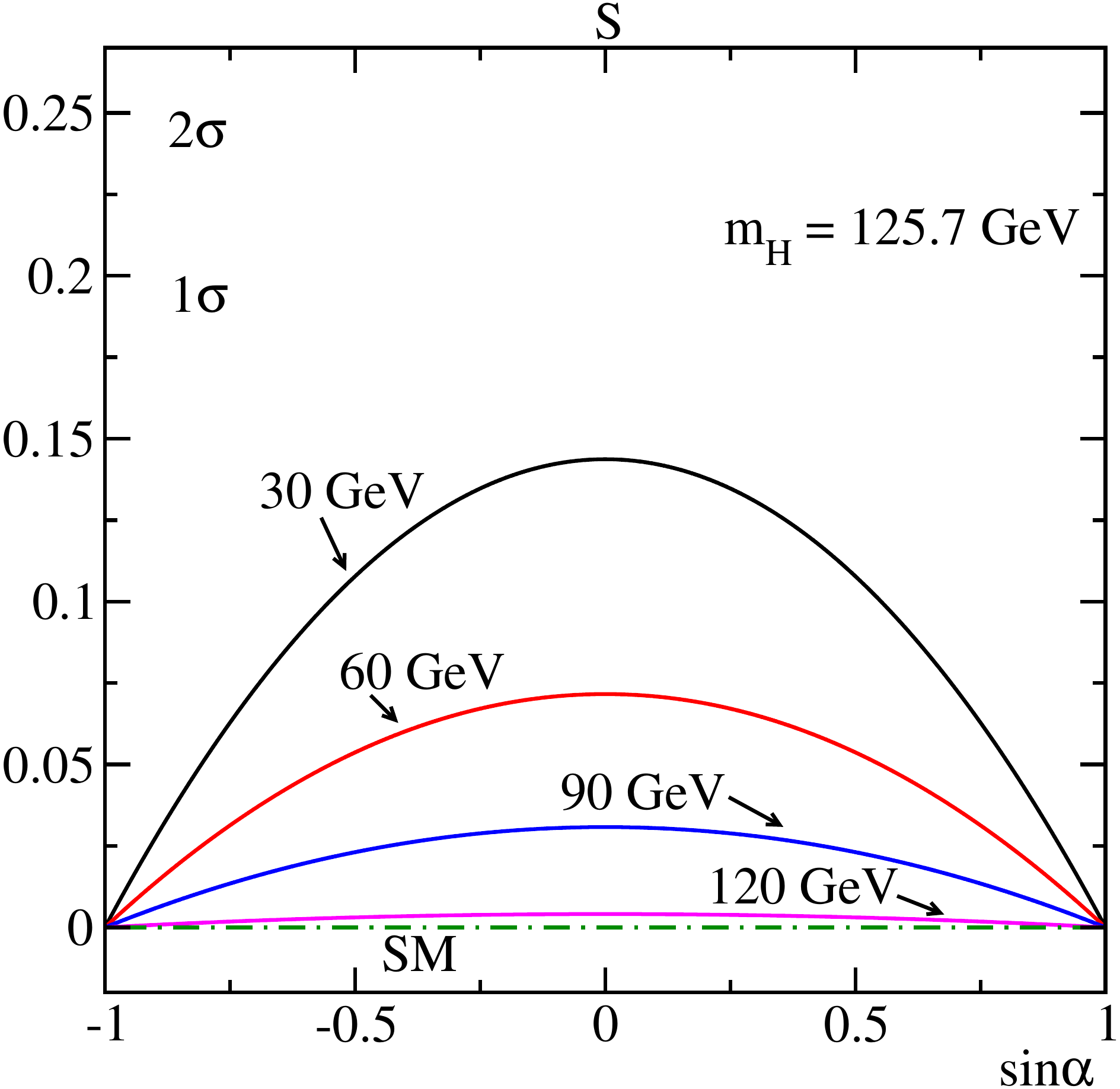} & \includegraphics[width=0.28\textwidth]{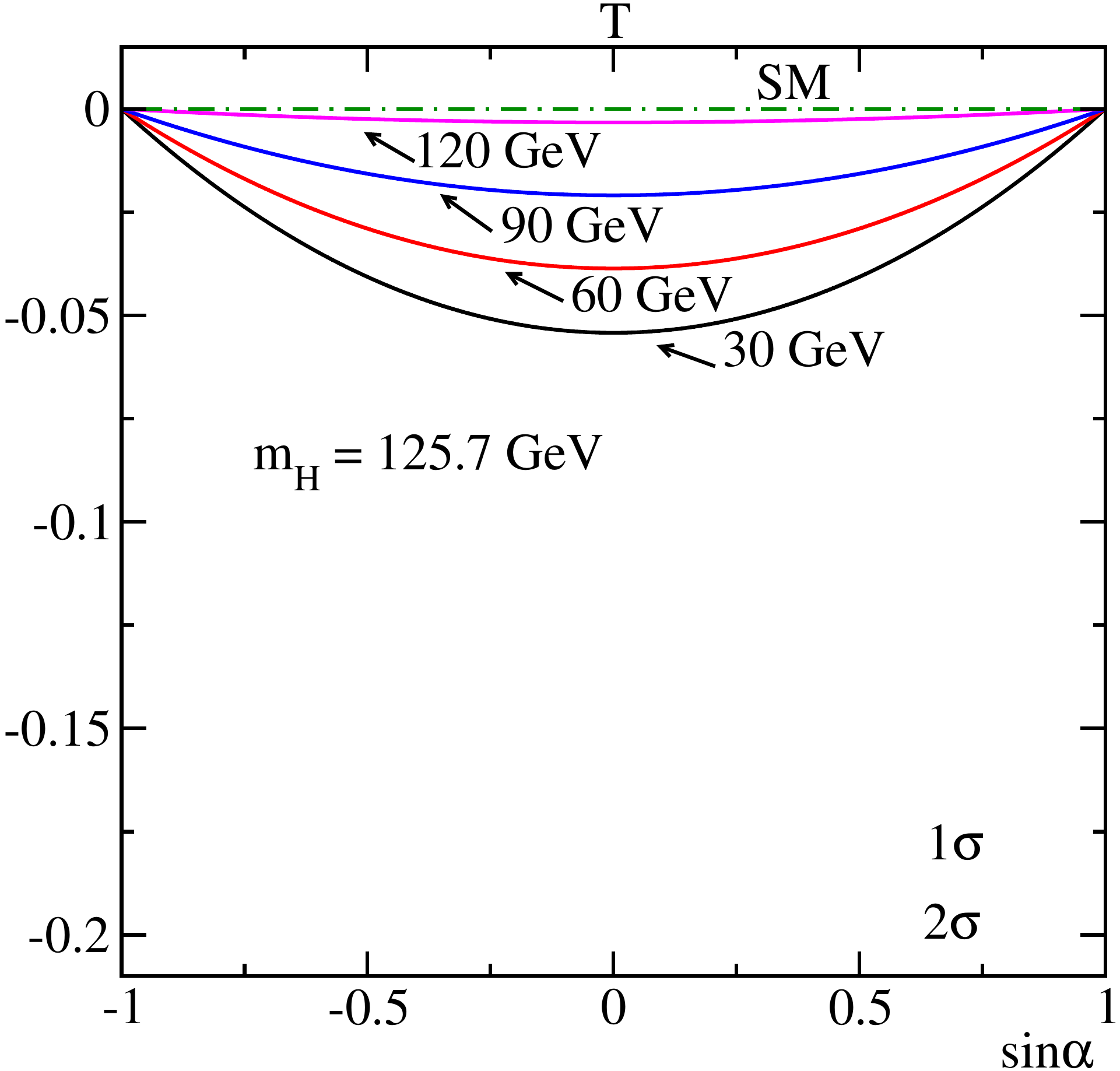} &  \includegraphics[width=0.28\textwidth]{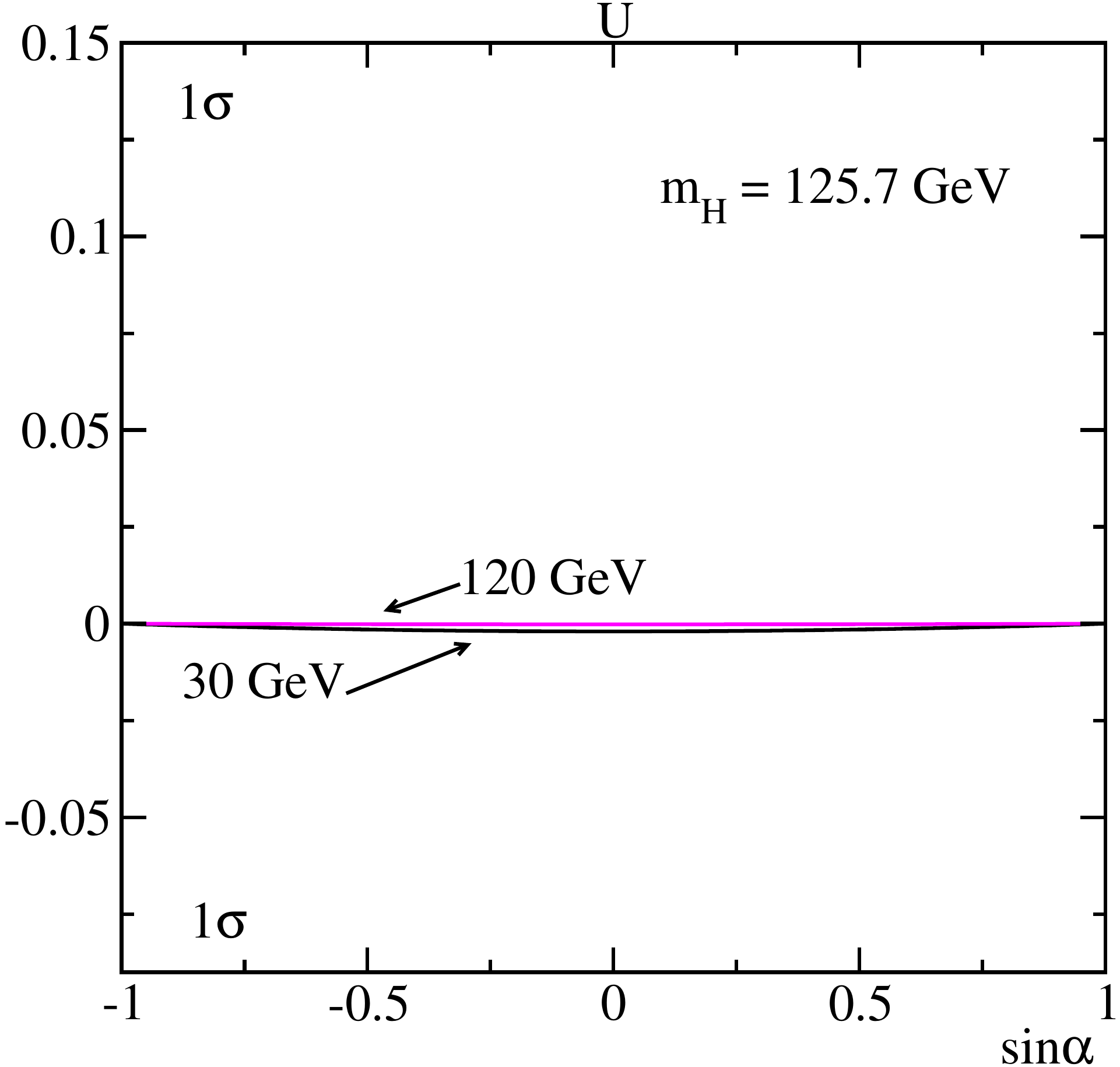} 
 \end{tabular}
 \end{center}
 \caption{Singlet model contributions
 to the oblique parameters $S$ (left), $T$ (center) and $U$ (right) as a 
 function of the mixing angle {for} representative
 heavy (upper row) and light (lower row) Higgs companion masses. 
 The dashed--dotted line represents the fiducial
 SM reference value $S,T,U = 0$.}
 \label{fig:oblique}
\end{figure}

\begin{figure}[t!]
\begin{center}
\includegraphics[width=0.5\textwidth]{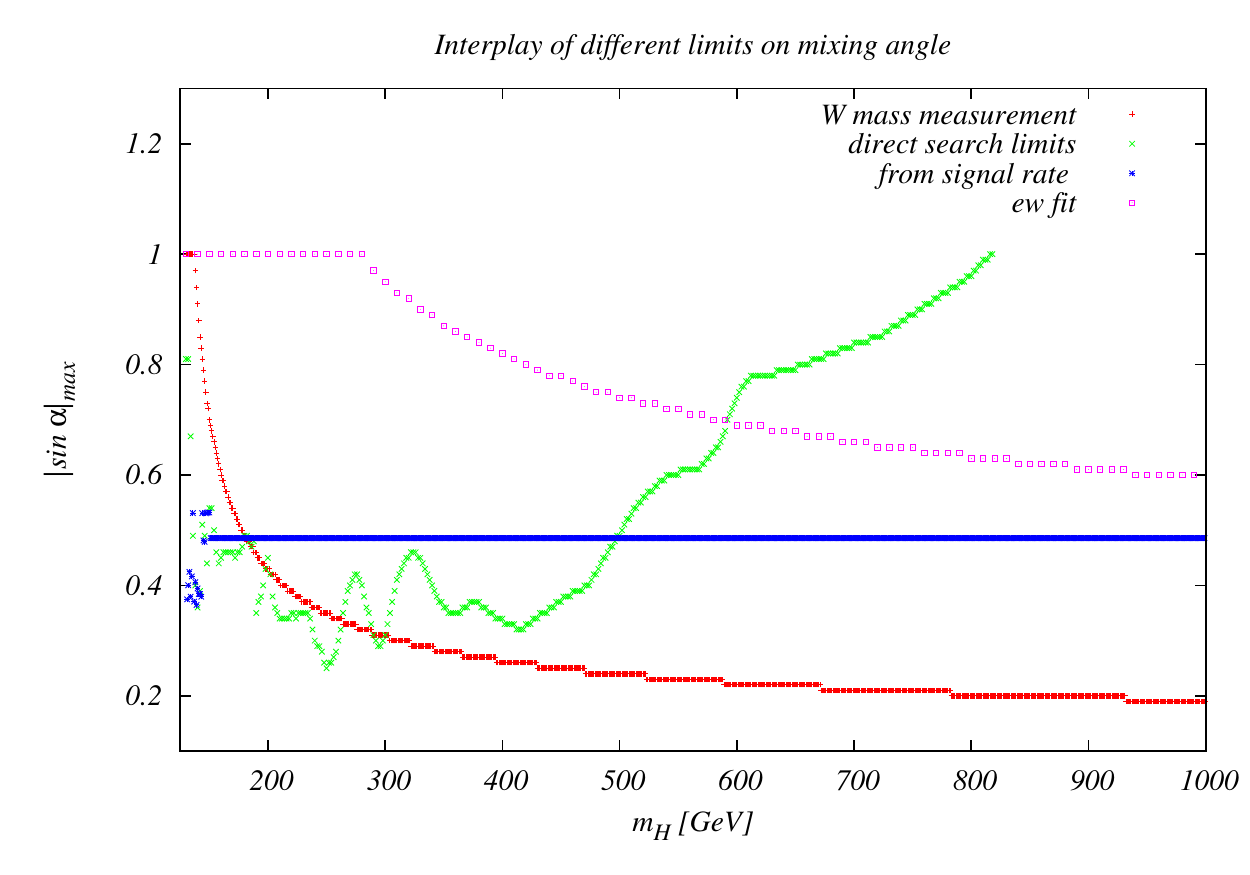}
\caption{\label{fig:expbounds} 
Upper limits on the mixing angle $|\sin\alpha|_{\text{max}}$ from i) the $m^{\text{exp}}_W$ measurement; ii) direct collider searches; 
 iii) compatibility with the $m_{\hzero} = 125.7$ GeV LHC Higgs
signal strength measurements, and iv) electroweak precision tests using $S,\,T,\,U$.} 
\end{center}
\end{figure}

Next, we also consider the constraints to the maximal values of the mixing angle 
{stemming from  direct collider searches  
and the averaged LHC Higgs signal strength measurements $[\bar{\mu}^{\text{exp}}]$. 
For the former, we use {\sc HiggsBounds} \cite{Bechtle:2013wla} which incorporates detailed information from {around 300} search channels from the LEP, Tevatron, and LHC experiments, to
extract upper {(resp. lower)} limits {on the mixing angle} {$|\sin\alpha|_{\text{max (min)}}$}   
 as a function of the {heavy {(resp. light)} Higgs companion}.  
In the mass range $m_{\Hzero} = 200 - 1000\,\GeV$, the primary collider
limits follow from 
the CMS four--lepton mode search \cite{CMS-PAS-HIG-13-002}; for lower masses {additional} channels are equally important \cite{CMS-PAS-HIG-12-045,CMS-PAS-HIG-13-003}.
Concerning 
the Higgs signal strength, we use the most recent values reported in \cite{ATLAS-CONF-2014-009,CMS-PAS-HIG-13-005}  
\begin{equation}
\mu_\text{ATLAS}\,=\,1.30\,\pm\,0.18,\,\mu_\text{CMS}\,=\,0.8{0}\,\pm\,0.14 \quad {\text{wherefrom}   \qquad \bar{\mu}^{\text{exp}}\,=\,1.05\,\pm\,0.11} \label{eq:hsfits}.
\end{equation}


A {word} of caution should be given here. 
Note that {these} best--fit estimates and C.L. limits
{are not tailored to any particular model.}
This means for instance that,  
although the singlet model {can only yield} {a suppressed Higgs signal strength} ${\mu}^{\text{sing}} \leq 1$,
such restriction is not enforced beforehand {when deriving the results in Eq.~\eqref{eq:hsfits}}. 
Dedicated model--specific analyses should {therefore include such model--dependent fit priors}, which 
{would eventually} modify the resulting bounds\footnote{We thank A. Straessner for clarifying comments regarding this point.}.

\smallskip{}
{To estimate the mixing angle range for which $\bar{\mu}^{\text{sing}}$ is compatible with
the LHC observations
$[\bar{\mu}^{\text{exp}}]$, we identify the light (heavy) singlet model mass--eigenstate with the SM Higgs boson
and assume a global rescaling $\bar{\mu}^{\text{sing}}/\bar{\mu}^{\text{SM}} \simeq \cos^2\alpha\,(\sin^2\alpha)$. 
{In this simple estimate, we do not entertain the possibility that two 
eigenstates with almost degenerate masses $m_{\hzero} \simeq m_{\Hzero}$ {could} contribute to the LHC Higgs signal}.
Allowing up to $2\sigma$--level deviations,
we obtain} upper (lower) mixing angle limits of
$|\sin\al|\,\leq\,0.42\,\lb |\sin\al|\,\geq\,0.91 \rb$. 
In the latter case, namely for $m_{\hzero} <  m_{\Hzero} = 125.7$ GeV, these are in fact 
comparable to or even stronger than the mass constraints from direct collider searches alone, {as well as from
the limits on the oblique $[S,T,U]$ parameters.
{This result implies} that 
the parameter space for the case of $m_{\Hzero}\,=\,125.7\,\GeV > m_{\hzero}$ is severely restricted. Consequently,
most regions in Figures~\ref{fig:oversa}-\ref{fig:overmass} 
for which quantum corrections would shrink the [$m_W^{\text{th}}-m_W^{\text{exp}}$] discrepancy below the $1\sigma$--level
are in practice precluded by the LHC signal strength measurements.
 {One should also bear in mind that, for very {light} $m_{\hzero}$ masses, additional constraints {from}
low--energy observables may play a significant role, see e.g. \cite{Isidori:2013cla,Gonzalez-Alonso:2014rla,Isidori:2014rba} and references therein.} 

\medskip{}
On the other hand, 
larger regions of parameter space are still allowed when
$m_{\hzero}\,=\,125.7\,\GeV<m_{\Hzero}$.
For this case, compared vistas of the different model constraints  
are displayed in Figure \ref{fig:expbounds}. 
We sweep the heavy Higgs masses in the range $130 - 1000$ GeV
and overlay the upper {bounds} on the mixing angle $|\sin\alpha|_{\text{max}}$
from each constraint individually.
While direct search bounds and signal strenght measurements dominate in the low--mass region,
both are superseded by the W--boson mass measurement [$m_W^{\text{exp}}$] 
for $m_{\Hzero}\,\gtrsim\,300\,\GeV$. This can once again be attributed to the 
Higgs--mediated corrections encoded within $\Delta r$, 
which increase with $m_{\Hzero}$ and are ultimately linked to the custodial symmetry breaking.
In turn, the limits {imposed by} the {correlated} oblique $[S,T,U]$ parameters {(cf. the magenta curve
in Fig.~\ref{fig:expbounds}) are also milder than those obtained from the 
[$m_W^{\text{sing}} - m_W^{\text{exp}}$] comparison. 
This result is after all not surprising (cf. e.g. Ref~\cite{Flacher:2008zq}) and reflects
the fact in a global fit (in this case {parametrized by} $[S,T,U]$),
the effect of the individual observables involved in it can
balance each other in part. {The resulting C.L. limits are then smeared} with respect to 
{the situation in which} we separately consider the more constraining measurements (in our case {$m^{\text{exp}}_W$}) individually. 
In this regard, let us recall the {very accurate precision} (viz. $0.02\%$ level) available for
the W--boson mass measurement.}
We conclude that these
different sources of constraints are highly complementary {to each other} in the different
{heavy Higgs} mass regions, and in all
cases rule out substantial deviations from the SM--like limit, viz. 
mixing angles of $|\sin\alpha| \gtrsim 0.2 - 0.4$.\\

\section{Summary}
\label{sec:summary}

We have reported on the computation of the electroweak precision parameter $\Delta r$,
along with the theoretical prediction of the W--boson mass, 
{in the presence of one additional real scalar $SU(2)_L\otimes U(1)_Y$ singlet.}
The $\Delta r$ parameter trades the relation between the electroweak
gauge boson masses, the Fermi constant and the muon lifetime.
Its precise theoretical knowledge 
plays a salient role in the quest for physics beyond the SM.
The reason is twofold:
first, because $\Delta r$ and $m_W$ constitute a
probe of electroweak quantum effects and are therefore
sensitive to, and able to constrain, extended Higgs sectors; and second, 
due to the 
current $1\sigma$ discrepancy $|m^{\text{SM}}_W - m_W^{\text{exp}}| \sim 20$ MeV which,
if eventually growing with the more accurate 
upcoming {W--boson mass} measurements,  {it}
could become a smoking gun for new physics. 

\medskip{}
{In this work} we have combined 
the state--of--the--art SM prediction (available up to leading three--loop
accuracy) 
with the one--loop evaluation
of the genuine singlet model effects. 
The two possible realizations of the singlet--extended SM Higgs sector,
viz. {featuring} a {heavy or a light}
Higgs companion, have been separately examined. 
{Finally, we have confronted the constraints on the parameter space
stemming from [$m_W^{\text{exp}}$] to the limits 
imposed by i) direct collider searches; ii) Higgs signal strenght
measurements; and iii) the {bounds on [$S,T,U$] based 
on global fits to electroweak precision data.}

Our conclusions may be outlined as follows:

\begin{itemize}
 \item {The singlet model contributions to $\Delta r$ {and $m_W$}
 are characterized by: i) a global rescaling factor which depends on
 the mixing {between the two scalar mass--eigenstates}
 and reflects the universal suppression of 
 all Higgs boson couplings in this model; 
 ii) the additional
 exchange of the second Higgs boson, which {exhibits}
 a 
 logarithmic {screening--like} non-decoupling dependence with the Higgs mass.
 }
 \item{The singlet--induced {new physics} 
 effects may typically yield up to $\mathcal{O}(10)\%$ deviations
 in the $\Delta r$ parameter with respect to the SM prediction. 
 Due to the characteristic dependence on the Higgs masses,
 these departures are bound to be positive if the lightest mass eigenstate
 is identified with the SM Higgs boson. 
 Such a shift $\drsing > \drsm$ implies $|m_W^{\text{sing}} - m_W^{\text{SM}}| \sim 1 - 70$ MeV 
 with $m_W^{\text{sing}} < m_W^{\text{SM}}$,  
 which raises the tension with the current [$m_{W}^{\text{exp}}$] measurement. 
 These trends are reverted if we exchange the roles
 of the two mass--eigenstates and consider instead 
 a light Higgs companion with $m_{\hzero} < m_{\Hzero} = 125.7$ GeV. In that case 
 we retrieve $\drsing < \drsm$ and hence $m_W^{\text{sing}} > m_W^{\text{SM}}$, which 
 makes in principle possible to satisfy $m_W^{\text{sing}} \simeq m_W^{\text{exp}}$. The viability of these
 scenarios is nonetheless limited in practice, {as they are hardly compatible with}  the Higgs {signal strength} measurements.}
 \item{Tight upper bounds on the mixing angle parameter $|\sin\alpha|_{\text{max}}$ can be derived  
 when confronting [$m_{W}^{\text{sing}}$] to  [$m_W^{\text{exp}}$].  
These are particularly {stringent} for $m_{\hzero}\,=\,125.7\,\GeV$ and $m_{\Hzero}\,\gtrsim\,300\,\GeV$,
{and reflect the enhanced breaking of the (approximate) custodial symmetry of the SM.}
In fact, in this mass range they dominate over the additional} model constraints 
from direct collider searches and Higgs signal strength measurements, {as well as from
global electroweak {fits} traded by the oblique parameters $[S,T,U]$ }.
\end{itemize}

\bigskip{}
With the calculation of the $\Delta r$ parameter, 
we have taken one step towards a complete characterization 
of the one--loop electroweak effects
in the singlet extension of the SM. The knowledge
of $\Delta r$ is a key element in the evaluation of the 
electroweak quantum corrections to the Higgs boson
decays. Work in this direction is underway~\cite{prep}.

\bigskip{}
 
\section*{Acknowledgements}
TR would like to thank S. Abel, C. Pietsch, G.M.Pruna, H. Rzehak, T. Stefaniak, A. Straessner and D. St\"ockinger 
for useful discussions in relation 
to the work presented here. Part of this work has been done during the Workshop "After the Discovery: 
Hunting for a Non--Standard Higgs Sector" at the "Centro de Ciencias de Benasque Pedro Pascual". DLV 
is indebted to J. Sol\`a for the earlier common work and
the always enlightening discussions on these topics. DLV also wishes to  
acknowledge the support of the   
F.R.S.-FNRS ``Fonds de la Recherche Scientifique'' (Belgium). \\

\newcommand{\JHEP}[3]{ {JHEP} {#1} (#2)  {#3}}
\newcommand{\NPB}[3]{{\sl Nucl. Phys. } {\bf B#1} (#2)  {#3}}
\newcommand{\NPPS}[3]{{\sl Nucl. Phys. Proc. Supp. } {\bf #1} (#2)  {#3}}
\newcommand{\PRD}[3]{{\sl Phys. Rev. } {\bf D#1} (#2)   {#3}}
\newcommand{\PLB}[3]{{\sl Phys. Lett. } {\bf B#1} (#2)  {#3}}
\newcommand{\PL}[3]{{\sl Phys. Lett. } {#1} (#2)  {#3}}
\newcommand{\EPJ}[3]{{\sl Eur. Phys. J } {\bf C#1} (#2)  {#3}}
\newcommand{\PR}[3]{{\sl Phys. Rep } {\bf #1} (#2)  {#3}}
\newcommand{\RMP}[3]{{\sl Rev. Mod. Phys. } {\bf #1} (#2)  {#3}}
\newcommand{\IJMP}[3]{{\sl Int. J. of Mod. Phys. } {\bf #1} (#2)  {#3}}
\newcommand{\PRL}[3]{{\sl Phys. Rev. Lett. } {\bf #1} (#2) {#3}}
\newcommand{\ZFP}[3]{{\sl Zeitsch. f. Physik } {\bf C#1} (#2)  {#3}}
\newcommand{\MPLA}[3]{{\sl Mod. Phys. Lett. } {\bf A#1} (#2) {#3}}
\newcommand{\JPG}[3]{{\sl J. Phys.} {\bf G#1} (#2)  {#3}}
\newcommand{\FP}[3]{{\sl Fortsch. Phys.} {\bf G#1} (#2)  {#3}}
\newcommand{\PTP}[3]{{\sl Prog. Theor. Phys. Suppl.} {\bf G#1} (#2)  {#3}}

\bibliography{hnlo}
\end{document}